\documentclass[12pt]{article}
\usepackage{epsf}

\usepackage{epsfig,graphics}
\usepackage {graphicx}
\usepackage {epsfig}
\usepackage {subfigure}
\usepackage {tabularx} 
\usepackage{rotate}	
\usepackage{slashed}
\usepackage{xcolor}
\usepackage{cancel}

\usepackage{amsmath}
\usepackage{amsfonts}
\usepackage{amssymb}
\usepackage{graphicx}
\usepackage{cite}

\usepackage{fancyhdr}
\usepackage{hyperref}

\newcommand{\bmat}{\left(\begin{array}}
	\newcommand{\emat}{\end{array}\right)}

\def\yzero{\smash{\hbox{$y\kern-4pt\raise1pt\hbox{${}^\circ$}$}}}

\def\beq{\begin{equation}}
\def\eeq{\end{equation}}
\def\beqa{\begin{eqnarray}}
\def\eeqa{\end{eqnarray}}

\def\-{\hphantom{-}}

\def\s2{\frac{1}{\sqrt2}}

\def\beq{\begin{equation}}
\def\eeq{\end{equation}}
\def\beqa{\begin{eqnarray}}
\def\eeqa{\end{eqnarray}}

\def\IF{\relax{\rm I\kern-.18em F}}
\def\II{\relax{\rm I\kern-.18em I}}
\def\IP{\relax{\rm I\kern-.18em P}}
\def\IC{\relax\hbox{\kern.25em$\inbar\kern-.3em{\rm C}$}}
\def\IR{\relax{\rm I\kern-.18em R}}

\def\cn{{\cal N}}

\def\cp{{\cal P}}

\def\cv{{\cal V}}

\def\Dsl{\,\raise.15ex\hbox{/}\mkern-13.5mu D} 
\def\IZ{Z\kern-.4em  Z}

\def\yzero{\smash{\hbox{$y\kern-4pt\raise1pt\hbox{${}^\circ$}$}}}

\def\beq{\begin{equation}}
\def\eeq{\end{equation}}
\def\beqa{\begin{eqnarray}}
\def\eeqa{\end{eqnarray}}

\def\-{\hphantom{-}}

\def\s2{\frac{1}{\sqrt2}}

\def\beq{\begin{equation}}
\def\eeq{\end{equation}}
\def\beqa{\begin{eqnarray}}
\def\eeqa{\end{eqnarray}}

\def\IF{\relax{\rm I\kern-.18em F}}
\def\II{\relax{\rm I\kern-.18em I}}
\def\IP{\relax{\rm I\kern-.18em P}}
\def\IC{\relax\hbox{\kern.25em$\inbar\kern-.3em{\rm C}$}}
\def\IR{\relax{\rm I\kern-.18em R}}

\def\cN{{\mathcal N}}

\def\cp{{\mathcal P}}

\def\cv{{\mathcal V}}

\def\Dsl{\,\raise.15ex\hbox{/}\mkern-13.5mu D} 
\def\IZ{{\mathbb Z}}

\def\mg{{m_{3/2}}}

\def\mgd{{M^\delta_{3/2}}}
\def\mp{{M_{\text{P}}}}
\def\ms{{M_{\text{s}}}}
\def\mkk{{M_{KK}}}
\def\mkkt{{M^2_{\text{KK}}}}

\def\mpt{{M^2_{\text{P}}}}
\def\mpf{{M^4_{\text{P}}}}

\def\cp#1{\relax\ifmmode {\IP\kern-2pt{}_{#1}}\else $\IP\kern-2pt{}_{#1}$\fi}

\def\Mt{M_{\mathrm{tower}}}
\def\m32{m_{3/2}}
\def\mgd{m_{3/2}^\delta}
\def\tmem{T_{\text{mem}}}
\def\hn{\hat{n}}
\def\hr{\hat{r}}

\catcode`\@=11   
\newdimen\@rotdimen
\newbox\@rotbox  

\def\@vspec#1{\special{ps:#1}}
\def\@rotstart#1{\@vspec{gsave currentpoint currentpoint translate
		#1 neg exch neg exch translate}}
\def\@rotfinish{\@vspec{currentpoint grestore moveto}}
\def\@rotr#1{\@rotdimen=\ht#1\advance\@rotdimen by\dp#1%
	\hbox to\@rotdimen{\hskip\ht#1\vbox to\wd#1{\@rotstart{90 rotate}%
			\box#1\vss}\hss}\@rotfinish}
\def\@rotl#1{\@rotdimen=\ht#1\advance\@rotdimen by\dp#1%
	\hbox to\@rotdimen{\vbox to\wd#1{\vskip\wd#1\@rotstart{270 rotate}%
			\box#1\vss}\hss}\@rotfinish}%
\def\@rotu#1{\@rotdimen=\ht#1\advance\@rotdimen by\dp#1%
	\hbox to\wd#1{\hskip\wd#1\vbox to\@rotdimen{\vskip\@rotdimen
			\@rotstart{-1 dup scale}\box#1\vss}\hss}\@rotfinish}%
\def\@rotf#1{\hbox to\wd#1{\hskip\wd#1\@rotstart{-1 1 scale}%
		\box#1\hss}\@rotfinish}%
\def\rotate{\@ifnextchar[{\@rotate}{\@rotate[l]}}
\def\@rotate[#1]#2{\setbox\@rotbox=\hbox{#2}\@nameuse{@rot#1}\@rotbox}

\catcode`\@=12

\topmargin
-1.5cm
\textwidth
15.5cm
\textheight
23.5cm
\oddsidemargin
0.7cm
\evensidemargin
0.7cm

\begin{document}
	\makeatletter
	\@addtoreset{equation}{section}
	\makeatother
	\renewcommand{\theequation}{\thesection.\arabic{equation}}
	\pagestyle{empty}
	\vspace{-0.2cm}
	\rightline{ IFT-UAM/CSIC-21-40}
	\vspace{1.2cm}
	\begin{center}
		
		
		\LARGE{ A  Gravitino Distance Conjecture \\
			[13mm]}
		
		\large{ A. Castellano$^1$, A. Font$^{2,3}$, A. Herr\'aez$^4$ and L.E. Ib\'a\~nez$^1$
			\\[6mm]}
		\small{
			$^1$ {\em Departamento de F\'{\i}sica Te\'orica
				and Instituto de F\'{\i}sica Te\'orica UAM/CSIC,\\[-0.3em]
				Universidad Aut\'onoma de Madrid,
				Cantoblanco, 28049 Madrid, Spain}  \\[0pt]
			$^2${\em Facultad de Ciencias, Universidad Central de Venezuela, A.P.20513, \\[-0.3em]
				Caracas 1020-A,  Venezuela} \\[0pt]
			$^3$ {\em Max-Planck-Institut f\"ur Gravitationsphysik, Albert-Einstein-Institut\\ [-0.3em]
				14476 Golm, Germany}\\[0pt]
			$^4$  {\em Institut de Physique Th\'eorique, Universit\'e Paris Saclay, CEA, CNRS\\ [-0.3em]
				Orme des Merisiers, 91191 Gif-sur-Yvette CEDEX, France} 
			\\[5mm]
		}
		\small{\bf Abstract} \\[6mm]
	\end{center}
	\begin{center}
		\begin{minipage}[h]{15.22cm}
			We conjecture that in a consistent supergravity theory with non-vanishing gravitino mass, the limit $m_{3/2}\rightarrow 0$ is at infinite distance. In particular one can write \mbox{$\Mt \sim m_{3/2}^\delta$} so that as the gravitino mass goes to zero, a tower of KK states as well as emergent strings becomes tensionless. This conjecture may be motivated from the Weak Gravity Conjecture as applied to strings and membranes and implies in turn the AdS Distance Conjecture. We test this proposal in classical 4d type IIA orientifold vacua in which one obtains a range of values $\tfrac13 \le \delta \le 1$. The parameter  $\delta$ is related to the scale decoupling exponent in AdS vacua and to the $\alpha$ exponent in the Swampland Distance Conjecture for the type IIA complex structure. We present a general analysis of the gravitino mass in the limits of moduli space in terms of limiting Mixed Hodge Structures and study in some detail the case of two-moduli F-theory settings. Moreover, we obtain general lower bounds $\delta\, \geq \, \frac{1}{3}, \, \frac{1}{4}$ for Calabi--Yau threefolds and fourfolds, respectively. The conjecture has important phenomenological implications. In particular we argue that low-energy supersymmetry of order 1 TeV is only obtained if there is a tower of KK states at an intermediate  scale, of order $10^8$ GeV. One also has an upper bound for the Hubble constant upon inflation $H\lesssim m_{3/2}^\delta M^{(1-\delta)}_{\text{P}}$.
			
		\end{minipage}
	\end{center}
	\newpage
	\setcounter{page}{1}
	\pagestyle{plain}
	\renewcommand{\thefootnote}{\arabic{footnote}}
	\setcounter{footnote}{0}
	

	
	\tableofcontents
	
	\section{Introduction}

	In recent years important efforts have been made in order to identify the crucial properties of 
	the effective field theories (EFT) which may be UV-completed into a consistent theory of Quantum Gravity (QG).
	This is the purpose of the {\it Swampland Program}\cite{swampland,WGC,review,vafafederico,vanBeest:2021lhn},
	whose objective is to extract general properties which may be tested in terms
	of string theory or well understood semiclassical properties of black holes.   These general properties are often stated in terms of
	{\it Swampland Conjectures} which try to capture essential properties coming from QG.
	One of the most studied Swampland conjectures  is the Swampland Distance Conjecture (SDC)\cite{SDC}.
	It states that for large moduli, an infinite tower of states becomes massless at an exponential rate $m\sim e^{-\alpha d}$,
	with \emph{d} the traversed geodesic distance.
	In some ${\cal N}=2$ examples this tower has been shown to be populated by 
	charged states saturating BPS bounds which become massless \cite{Grimm1,Grimm2,Corvilain,Grimm:2019wtx,Gendler:2020dfp,Bastian:2020egp}.
	The value of the constant $\alpha$ has been computed in  different ${\cal N}=2$ systems and has been shown to be related to the 
	corresponding charges of these towers of states.
	Certain ranges for the possible values of $\alpha$ have been found \cite{Andriot:2020lea,Gendler:2020dfp,Bastian:2020egp}.
	In some limits it is strings which 
	become tensionless as an infinite distance limit is approached \cite{timo1,timo2,timo3,timo4,timoemergent,fhi,infinitons1,infinitons2,lmmv,lmmv2}.
	
	A seemingly different statement is given by  the AdS distance conjecture (ADC) \cite{lpv}.
	This applies to theories in which changing (flux) parameters one can obtain families of AdS vacua with vacuum energy going to zero,
	$|\Lambda|\rightarrow 0$.  The statement is that an infinite tower of massless states must appear in this limit,
	with $m\sim |\Lambda|^\lambda$.  In a strong version  $\lambda $  was conjectured to be $1/2$ for supersymmetric theories and larger for
	non-supersymmetric theories. In many AdS string theory examples the existence of these massless towers has been checked
	\cite{lpv,Gautason:2018gln,Blumenhagen:2019vgj,Font:2019uva,Buratti:2020kda}.
	The situation concerning the allowed values of $\lambda$ is less clear, with some existing counterexamples.
	
	The connection between the SDC and towers of charged states allows for an understanding of the very existence of that tower.
	For large moduli some gauge symmetry becomes global, and to avoid conflict with QG, a tower of charged states appears. 
	This is in agreement with the magnetic Weak Gravity Conjecture (WGC) \cite{WGC}.
	In the case of the AdS distance conjecture the situation is less clear. What precisely goes wrong when $\Lambda\rightarrow 0$? 
	No obvious direct  connection with a WGC argument is apparent in this case. Also, as already mentioned, it is not yet clear what the range of possible values for
	the exponent might be.
	More generally, a disappointing feature of both distance conjectures is that, although they may provide interesting 
	tests of Swampland ideas, they do not seem to give us information on the EFT which could be used 
	in phenomenological applications.
	
	From the point of view of the SDC or the ADC there is no particular direction in moduli space which should be  studied more carefully than any other. Any infinite limit may be equally interesting to test
	the conjectures.
	In this paper we argue that there is a natural limit which plays a special role, namely the limit in which the gravitino 
	mass goes to zero. There are several  reasons for this to be the case. Unlike other possible asymptotic field directions, 
	this limit is associated to a unique physical particle, the gravitino, which is the (super)partner of the graviton. 
	Only along certain field limits the gravitino may become massless and hence the existence of a gravitino naturally selects
	particular classes directions in the space of moduli.
	Furthermore,
	the $m_{3/2}\rightarrow 0$ limit corresponds to theories with small supersymmetry breaking which may have direct phenomenological 
	implications both for particle physics and cosmology. Thus the possibility of having low-scale supersymmetry breaking rests on the
	existence of such moduli directions and whether  an EFT in which the gravitino scale is well separated from the UV scales exists.
	
	In field theory one can have models with gravitino masses as small as desired with no apparent contradiction.
	In this paper we make the conjecture that this is not the case in an EFT coming from a consistent theory of QG.
	In particular we argue that the limit $m_{3/2}\rightarrow 0$ in quantum gravity is at
	infinite distance. We thus formulate a
	
	\leftline{\it Gravitino Distance Conjecture (GDC):}
	
	{ \it In a theory of quantum gravity, in the limit $m_{3/2}\rightarrow 0$,  an infinite tower of massless states appears	with lightest mass scale }
	\beq
	m\ \sim \ m_{3/2}^\delta  \ .
	\label{GDC}
	\eeq
	Hence for $\delta>0$ the scale of the gravitino cannot be arbitrarily separated from the UV scale. The fact that there could be a tower of light states in QG associated with such a fermionic field has been suggested in \cite{Palti:2020tsy}.
	Note that in the case of supersymmetric AdS vacua this directly implies the AdS distance conjecture, since 
	in that case $\Lambda=-3m_{3/2}^2$. However we propose this conjecture to be true also in non-supersymmetric AdS, Minkowski and dS vacua.
	Although all our discussion will concentrate in 4d theories, we expect the conjecture to be true also in higher dimensions in the presence of a 
	massive gravitino.  In the main text we will test this conjecture using as a laboratory 4d ${\cal N}=1$ theories, from type II compactified on orientifolds of Calabi--Yau threefolds (CY$_3$), as well as from Calabi--Yau fourfold (CY$_4$) compactifications of F-theory.\footnote{In the context of heterotic compactifications, the existence of a tower of gravitinos becoming light in the vanishing gravitino mass limit was explored in \cite{Antoniadis:1988jn}.}
	
	A first question with this proposal  is precisely  what goes wrong when the gravitino mass goes to zero, forcing a tower of states to appear.
	We find that,  in the context of flux type II orientifold vacua, in such a limit a set of strings and membranes become tensionless and
	the corresponding gauge coupling of the forms and 3-forms goes to zero as well, in conflict with general Swampland ideas. In particular, 
	axionic shift symmetries become continuous. Moreover, in the $m_{3/2}\rightarrow 0$ limit the KK scale is in general lighter than the scale of those 
	extended objects and it is this scale which becomes lightest first. Thus in eq.(\ref{GDC}) $m$ will typically be the
	KK scale in specific examples.

	One can also see  that the gravitino mass in general type IIA(B)  $\mathcal{N}=1$ orientifold AdS and Minkowski minima 
	depends only on the complex structure (K\"ahler and dilaton) moduli. 
	Thus in  limits  of large 
	complex structure (in IIA)  a gravitino becomes massless and in general earlier than any 
	other state in the theory. Thus one expects that  the tower of massless states associated to the gravitino will be related 
	to the towers of states becoming light (according to the SDC) at large complex structure. Still, large K\"ahler moduli are also
	in general required to remain within the perturbative regime in a consistent EFT.
	The parameters $\delta$ for the gravitino and $\alpha$  for the SDC are thus expected 
	to be related in general.
	
	For general CY$_3$ orientifolds one finds a lower bound $\delta>1/3$. To gain more information we use 
	a class of   $\mathbb{Z}_2\times \mathbb{Z}'_2$ type IIA   toroidal orientifolds with fluxes  as a laboratory and explore the relationship  between  the gravitino mass
	and the KK-towers along different directions in the space of moduli, i.e. the range of values for $\delta$.
	In the case of AdS vacua that means along directions determined by the fluxes whereas in 
	Minkowski we deal with no-scale vacua in which some of the moduli are undetermined. One finds values in  the range $\tfrac23 \le \delta \le 1$. 
	Since $m_{3/2}\rightarrow 0$ limits happen for large complex structure, one can directly connect the values of $\alpha$ 
	in the SDC of the complex structure to the values of $\delta$. We observe that the range of $\delta$ we find matches with the range 
	of values for  $\alpha$ found  in the literature for ${\cal N}=2$ systems \cite{Gendler:2020dfp,Bastian:2020egp,Andriot:2020lea}.
	In the case of examples of dS runaway potentials, there is no minima but one can study the
	behavior of the field dependent gravitino mass, $m_{3/2}=e^{K/2}|W|$, which now may depend on all 
	complex structure and K\"ahler moduli. The connection between the  gravitino mass and the tension of membranes is 
	thus very transparent and,  as the gravitino mass goes to zero, so does the tension of the membranes.
	Still for large moduli a tower of KK states also becomes massless. The potentials one finds are consistent with the
	asymptotic dS conjecture.
	
	Some more general properties for any CY orientifolds may be studied using the formalism of limiting Hodge structures
	\cite{Grimm1, Grimm2, Corvilain,Grimm:2019wtx,Gendler:2020dfp,Grimm:2019bey,Grimm:2019ixq,Grimm:2020cda,Grimm:2021ikg,Bastian:2020egp,Grimm:2020ouv}.
	We present a general study of the $m_{3/2}\rightarrow 0$ limit in terms of that formulation within an F-theory general setting.
	Focusing on the 
	two-moduli class of examples, we generalize our results from type IIA orientifolds. Moreover,  we show general constraints for the exponents $\delta$ which are 
	associated to strings becoming tensionless as we approach the limit of vanishing gravitino mass. In particular one obtains again $\delta\geq 1/3$ for CY$_3$ orientifolds and
	$\delta\geq 1/4$ for F-theory flux compactifications on CY$_4$.  
	This is also consistent with the toroidal orientifold models systematically analysed. Furthermore, the precise relation between the GDC and the SDC for the case of Minkowski no-scale vacua is generalized using this mathematical machinery and explicit bounds for $\delta$ in terms of the SDC parameter, $\alpha$, 
	are given. Matching with the explicit models studied in this paper is also found.
	
	The possible values for $m_{3/2}$ have phenomenological implications.
	Depending on the value of $\delta$, the gravitino mass decouples from the KK tower or not.
	As we said, in AdS and no-scale Minkowski examples one generally finds $\tfrac13 \le \delta \le 1$ and maximum decoupling occurs for 
	$\delta = 1/3$, although for the leading KK towers analyzed in our explicit examples the minimum we find is $\delta=2/3$. 
	Phenomenologically interesting values for $m_{3/2}$ require, if the GDC holds, 
	values for the UV thresholds well below the standard unification scales.
	Thus, assuming e.g. $m_{3/2}\ =\ 1$ TeV, just around  LHC reach, it must be that $M_{KK}\sim 10^8$   GeV if $\delta=2/3$.
	Hence a large desert from the electro-weak to the unification scale   is not possible. If on the other hand we set  $m_{3/2}$ at an
	intermediate scale $m_{3/2}\sim 10^{10}$ GeV, as in certain classes of phenomenologically interesting models,  one obtains $M_{KK}=10^{13}$ GeV.
	Cosmology is also affected by these lower values of UV scales since the Hubble parameter is then forced to obey 
	$H\ \lesssim \ m_{3/2}^\delta \ M^{(1-\delta)}_{\text{P}}$.
	
	The structure of this paper is as follows. In the next section we review the SDC and the ADC,
	which are intimately connected with the GDC proposed here. In section \ref{sec:GDC} we define the GDC 
	and how it implies 
	the ADC. We also explain how in the $m_{3/2}\rightarrow 0$ limit tensionless membranes arise, giving a WGC interpretation to its singular 
	behavior. In section \ref{sec:iia} we present explicit classes of type IIA toroidal orientifolds in which the GDC is tested, indicating how in
	AdS and Minkowski vacua the $m_{3/2}\rightarrow 0$ limit is controlled by  the dilaton and complex structure moduli. 
	However, remaining in the perturbative regime generically requires also large K\"ahler moduli. 
	A general bound $\delta>1/3$ is obtained, although toroidal orientifolds are restricted to 
	$2/3<\delta <1$. We additionally show how the GDC and the SDC are connected and the exponent $\delta$ is proportional to 
	the exponent $\alpha$  in the SDC. We also describe how tensionless strings also appear in that limit. Section \ref{sec:limits} is 
	devoted to the study of the GDC for 
	general  CY$_3$ and F-theory compactifications, using the formalism of limiting Mixed Hodge Structures, which is also briefly introduced. 
	Constraints on the exponents $\delta$ are obtained for this general case, showing how the results obtained for toroidal orientifolds are
	generalized. The phenomenological implications of the GDC are studied in section \ref{sec:pheno}, where some implications on the scale of supersymmetry breaking 
	and cosmology are briefly discussed.
	We leave section \ref{sec:conclusions} for some final comments and conclusions.
	
	\vspace{1.0cm}
	Note: As we were submitting this paper to the arXives, the paper \cite{Cribiori:2021gbf} appeared with some partial overlap with this one.

	\section{The Swampland Distance Conjecture and the Anti-de Sitter Distance Conjecture}
	\label{sec:SDC}
	
	In order to set the stage, we review in this section the Swampland Distance Conjecture (SDC) \cite{SDC} and the Anti-de Sitter Distance Conjecture (ADC) \cite{lpv}, as they turn to be closely related to our Gravitino Distance Conjecture (GDC), that we present in the next section. Roughly speaking, both conjectures state that the regime of validity of any  Effective Field Theory (EFT) which arises as a low energy limit of Quantum Gravity (QG) is limited.  They predict that any particular EFT description breaks down by the appearance of a infinite tower of states that become light if we insist in describing configurations which are \textit{very far} from the original one.
	
	To be precise, consider a gravitational theory with a moduli space whose metric is given by the kinetic terms. The SDC then states that starting at a point $P$ in  moduli space, and moving towards a point $Q$ an infinite geodesic distance away from $P$, one encounters a tower of states whose masses (in Planck units) become exponentially light as 
	\begin{equation}
	\label{eq:SDC}
	\dfrac{ \Mt (Q) }{ \mp (Q) } \, \sim \,  \dfrac{ \Mt (P) }{ \mp (P) } \, e^{- \alpha d(P, \, Q)}\, ,
	\end{equation}
	where $\mp (\cdot )$ is the EFT Planck Mass at the point $\cdot$ in moduli space, and $d(P, \, Q)$ is the geodesic distance between the points $P$ and $Q$.
	
	The SDC is one of the most studied Swampland Conjectures and it has passed numerous non-trivial checks. It has been thoroughly studied at the limits of moduli space of type II compactifications on CY$_3$, where towers of light BPS states have been identified \cite{Grimm1, Grimm2, Corvilain}. Moreover, its relation with extended objects (and instanton corrections) becoming light at infinite distance points has been an essential ingredient since its original formulation and it has been thoroughly explored in \cite{timo1, timo2, timo3, infinitons1,  fhi, timo4,Grimm:2019wtx, timoemergent,infinitons2, Gendler:2020dfp,lmmv, Klaewer:2020lfg}. In fact,  tensionless strings seem to play a special role, as emphasized by the \textit{Emergent String Conjecture} \cite{timoemergent}, which states that any infinite distance point is either a decompactification limit or a limit in which a string becomes tensionless. More recently, the special role of strings has also been revisited in the language of 4d supersymmetric EFTs with the \textit{Distant Axionic String Conjecture} \cite{lmmv}, which states that all infinite distance points of a 4d EFT correspond to a tensionless axionic string.\footnote{It is interesting to note that whereas the \textit{Emergent String Conjecture} deals with strings in any number of dimensions, the relevant objects for the \textit{Distant Axionic String Conjecture} are codimension-two objects, which are strings only in four dimensions.}

	The ADC can be seen as a particular case of a generalized version of the SDC, namely the Generalized Distance Conjecture \cite{lpv}. The idea is to generalize the notion of distance in moduli space to a notion of distance applicable to any (tensor) field configuration, with a generalized metric given once again by the kinetic terms of the corresponding tensor fields. The claim is again that an infinite tower of states becomes light exponentially with the proper distance when the distance diverges. When applied to families of vacua with different values for the cosmological constant (c.c.), as explained in \cite{lpv}, this implies the existence of a tower of states becoming light when the c.c. goes to zero, according to
	\begin{equation}
	\label{eq:ADC}
	\dfrac{\Mt }{\mp} \, \sim \, \left| \Lambda \right|^\lambda \, ,
	\end{equation}
	with $\lambda$ a positive $\mathcal{O}(1)$ number. The strong version of the conjecture states that $\lambda=1/2$ for supersymmetric vacua and for the AdS case it implies the absence of scale separation between the AdS mass and the mass of a KK tower, which is the one responsible for the breakdown of the corresponding EFT. Moreover it has also been argued \cite{lpv} that $\lambda\geq 1/2$ for AdS and $\lambda \leq 1/2$ for dS.
	
	The  weak version of the ADC seems to be supported by all known examples. However, the strong version is in tension with e.g. the class of type IIA vacua found in \cite{dgkt, cfi}, where a family of supersymmetric vacua yields $\lambda=7/18$ and therefore exhibits scale separation. It has also been pointed out that the value 
	$\lambda=1/D$, with $D$ the number of dimensions, also naturally appears if one imposes  stability of the ADC under dimensional reduction
	\cite{Rudelius:2021oaz,Gonzalo:2021fma}.
	The models in \cite{dgkt, cfi} have recently been revisited from a 10d point of view \cite{Junghans:2020acz,Marchesano:2020qvg} (see also \cite{Blumenhagen:2019vgj,Font:2019uva,Buratti:2020kda} for related ideas about this issue) and no inconsistency has been found, but a full 10d solution is still missing and it would be required to clarify whether they are a robust counterexample to the strong ADC.
	
	It is therefore clear that even though the breakdown of  gravitational EFTs by the appearance of  light towers of states seems to be ubiquitous in String Theory, this generality makes it harder to pinpoint the towers that could be more relevant for connecting this to our Universe. In this regard, we will focus in this article on a particular limit, namely the one associated with the gravitino mass going to zero. This is particularly interesting for several reasons. First, the gravitino belongs to the gravity multiplet in supersymmetric EFTs and it is  therefore intrinsically tied to the gravitational character of the theory, as well as always present in all  supersymmetric EFTs. Second, the gravitino mass gives the scale of supersymmetry breaking in vacua with spontaneously broken supersymmetry, so it is an interesting quantity to look at from a phenomenological point of view. Third, the gravitino mass is typically related to the tension of a (codimension-one) membrane, and therefore its massless limit corresponds to a tensionless membrane limit. In this limit a generalized global symmetry would typically be restored and a tower of states is expected to prevent this from happening. In some sense, one could say that the Gravitino Distance Conjecture, that we present in the next section, allows us to unify the SDC and the ADC when the special limits selected by the vanishing of the gravitino mass are considered. In this sense, not only are we adding one conjecture to the already rich web of Swampland Conjectures, but also connecting it to some of the existing ones as well as recovering non-trivial results for the bounds on their parameters \cite{Gendler:2020dfp, Andriot:2020lea, Bastian:2020egp}, as we will explain later.

	\section{The Gravitino Distance Conjecture}
	\label{sec:GDC}

	Given the motivations just discussed we propose a  \textit{Gravitino Distance Conjecture} (GDC):
	
	{\it Consider a supersymmetric  theory with a non-vanishing gravitino mass $\m32$. In the limit $m_{3/2}\rightarrow 0$, a tower of states becomes light according to
		\beq 
		\label{eq:GDC}
		\dfrac{\Mt}{\mp} \, \sim \, \left(\dfrac{\m32}{\mp}\right)^\delta  \ , \ 0<\delta \leq 1 \, .
		\eeq
	}
	
	In the following we will present evidence for the GDC from different perspectives, as well as several connections to other Swampland Conjectures, but before going into that, some clarifications are in order. First of all, let us clarify that we are not claiming that vacua in which the gravitino is exactly massless belong to the Swampland. Instead, our claim is that one cannot continuously go from a non-vanishing gravitino mass to the limit $\m32\rightarrow 0$ without encountering an infinite tower of states, that is, the two configurations are at infinite distance from each other. Note that this is analogous to e.g. the claim in the ADC, which does not put Minkowski vacua in the Swampland but instead states that they cannot be approached smoothly from vacua with non-vanishing cosmological constant  without an infinite tower becoming massless. Second, since the gravitino mass typically depends on the moduli of the theory, we can think of approaching the limit $\m32 \rightarrow 0$ in two qualitatively different ways. On the one hand, when some of the moduli on which $\m32$ depends remain unfixed and their vevs can be freely adjusted to make the gravitino mass as small as one desires. 
	On the other hand, considering families of vacua where all the scalars on which the gravitino mass depends are fixed (e.g. by fluxes that source a potential), and by scanning the family (e.g. by changing the fluxes) one can make the gravitino mass go to zero. These two situations are reminiscent of the SDC and the ADC, as we will clarify later. Moreover one can also have a mixed situation in which the space of moduli consists of a discrete part, which includes those moduli that are fixed and can still vary with fluxes, and a continuous part, which includes the unfixed moduli (i.e. flat directions). We thus can define a {\it space of moduli  ${\cal M}_{sm}$} (rather than a moduli space) which has a direct product structure
	\beq
	{\cal M}_{sm} \ =\ {\cal M}_{discrete} \otimes \ {\cal M}_{continuous} \ .
	\eeq
	
	We will present examples of all these situations later on. In particular, in this work we focus on 4d $\cN =1$ compactifications for concreteness, but let us remark that we expect the GDC to be equally valid for higher dimensions or more supersymmetries. 
	For the moment, we recall a few relevant generalities about the gravitino mass in 4d $\cN=1$ supergravity. First, given a K\"ahler potential, $K$, and a superpotential, $W$, the mass of the gravitino (in Planck units) takes the form
	\begin{equation}
	\label{eq:m32}
	\m32\, = \, e^{K/2} \, |W|\, .
	\end{equation}
	In Minkowski vacua, a massive gravitino implies that supersymmetry is broken, and therefore $\Lambda_{\textrm{\cancel{SUSY}}} \sim \m32$. 
	In supersymmetric AdS vacua, the gravitino mass is directly related to the cosmological constant, which takes the value 
	(in Planck units) \mbox{$\Lambda = -3 e^K |W|^2=-3 \,m_{3/2}^2$}. In this case, supersymmetry may remain unbroken even if the gravitino gets a mass\footnote{This is due to the form of the supersymmetry algebra in an AdS background, see e.g.\cite{Freedman:2012zz}.}. When supersymmetry is broken the gravitino mass also sets the overall supersymmetry breaking scale in AdS and dS vacua, although such a precise equality does not apply.

	\subsection{Relation with the ADC}
	
	It is particularly relevant to examine the GDC in the case of supersymmetric AdS vacua, as in this setup it can be shown to be equivalent to the ADC. As explained above, considering an $\cN=1$ theory with an AdS supersymmetric vacuum yields
	\beq
	\label{eq:ccsusy}
	\Lambda\, =\, -3e^K|W|^2\ =\ -3 m^2_{3/2} \, ,
	\eeq
	where $K$ and $W$ are evaluated at the minimum. The GDC then implies 
	\beq
	\label{eq:GDC-ADC}
	\Mt \, \sim \,  \mgd \ \sim \ |\Lambda|^{\delta/2} \ .
	\eeq
	By comparing with eq. \eqref{eq:ADC}, the ADC can be stated saying that, as the gravitino mass goes to zero, necessarily a tower of states becomes massless, and we can identify the parameters in both conjectures as $\lambda=\delta /2$. Therefore, all supersymmetric examples that fulfill the (weak version of the) ADC are also in agreement with the GDC. In non-supersymmetric AdS vacua this identity between $\m32$ and the cosmological constant does not hold in general, but  given that the ADC is supposed to apply also to non-supersymmetric vacua it is reasonable to conjecture that also in this case 
	\beq
	\Mt \, \sim \,  \mgd \, .
	\eeq
	We therefore expect that as the gravitino mass goes to zero, there is a tower of states which becomes massless, also in the non-supersymmetric case. 
	
	In the next section we present several type II models in which both supersymmetric and non-supersymmetric vacua are examined, and all of them agree with our conjecture (with $1/3\leq \delta \leq 1$), even in the non-supersymmetric configurations. Furthermore, the general argument presented in the next subsection, based on 3-forms and membranes is equally valid in the non-supersymmetric vacua. Still, it  would be interesting to check more  examples of non-supersymmetric AdS.

	\subsection{Relation with the WGC}
	\label{ss:wgc}
	
	There is an interesting connection between the limit in which the gravitino mass vanishes and the existence of a membrane with vanishing tension. Consider a flux configuration, with gravitino mass given by eq. \eqref{eq:m32}. A BPS membrane interpolating between a Minkowski vacuum without fluxes, and the aforementioned flux configuration has a tension given by \cite{Gukov:1999ya,Taylor:1999ii} (see also \cite{Ceresole:2006iq, lmmv})
	\begin{equation}
	\label{eq:TmemBPS}
	\tmem \, = \, 2 e^{K/2}|W| \, ,
	\end{equation}
	in Planck units. In string compactifications one can interpret this membrane as a bound state of D$p$ or NS5 branes wrapping the appropriate cycles in the internal geometry. As a consequence, the limit in which the gravitino mass goes to zero is also the limit in which such a membrane becomes tensionless. Let us try to relate this to the WGC for membranes in 4d, which takes the form \cite{WGC16, lmmv}
	\begin{equation}
	\label{eq:WGC}
	M^{-2}_{\text{P}} \gamma_\mathrm{ext}^2 T^{2} \leq e^{2} Q^{2} \,
	\end{equation}
	where, $e$ is the gauge coupling associated to the 3-form that couples to the membrane, and $Q$ is its quantized charge. $\gamma_{\mathrm{ext}}$ is the charge-to-mass ratio of a extremal solution, that generically depends on the scalars of the theory. Note that for codimension-one objects, and due to the strong backreaction that they produce, the notion of extremality is subtler than for high-codimension states. However, we will take the approach in \cite{lmmv} and consider extremal membranes as infinite flat membranes, whereas superextremal ones are associated with bubbles that can mediate transitions. The F-term scalar potential generated by the fluxes sourced by the membrane can be dualized and expressed in terms of 3-forms, as shown for string compactifications in \cite{Bielleman:2015ina, HIMZ} and for $\mathcal{N}=1$ compactifications in \cite{Farakos:2017jme, Farakos:2017ocw, Bandos:2018gjp, Bandos:2019wgy,Lanza:2019xxg}, yielding the result 
	\begin{equation}
	\label{eq:V3forms}
	V=\frac{1}{2} Z^{A B} Q_{A} Q_{B},
	\end{equation}
	where the RHS is a generalization of the RHS of eq.\eqref{eq:WGC} for several 3-forms with kinetic term $\mathcal{S}_{ \mathrm{kin}}\sim \int Z_{A B} F^{A} \wedge \star F^{B}$, so that $Z^{AB}$ plays the role of the gauge coupling squared and $Q^A$ are the quantized charges of the membranes, which can be identified with the fluxes sourced by them. For $\cN=1$ compactifications the F-term scalar potential can also be written as
	\begin{equation}
	\label{eq:VFterm}
	V=e^{K}\left\{K^{I \bar{J}} D_{I} W \overline{\left(D_{J} W\right)}-3|W|^{2}\right\}
	\end{equation}
	with $K_{IJ}$ the K\"ahler metric, and $D_I=\partial_I + K_I$ the K\"ahler covariant derivative. These two ways of expressing the scalar potential can be interpreted as a no-force condition between membranes interpolating between a fluxless Minkowski vacuum and the one with the fluxes given by $Q^A$, with eq.\eqref{eq:V3forms} giving the electric interaction, and the two terms of eq.\eqref{eq:VFterm} describing the scalar and gravitational interactions respectively \cite{alvaro}. We can therefore identify the scalar potential generated by the membranes as the RHS of the WGC \eqref{eq:WGC}, and the LHS being proportional to the gravitational exchange given by the second term in the F-term scalar potential.
	
	Let us first consider BPS membranes interpolating between the fluxless Minkowski vacuum and a supersymmetric AdS one. This is the only interesting supersymmetric solution for us here as it is the only one that allows a non-vanishing gravitino mass. In this case, we have $D_I W=0$, and hence $V\, =\, -3 e^K |W|^2$. The physical charge of the membranes is then proportional to the gravitino mass, and in the limit in which it goes to zero so does the corresponding 3-form gauge coupling. Therefore we can see the gravitino mass going to zero as a consequence of the WGC for membranes, and we expect a tower of states becoming massless in that limit to prevent this higher-form symmetry from becoming global. In fact, this is nothing but the fact that for a BPS membrane its charge is proportional to its tension, and therefore as its tension goes to zero so does its charge, saturating the WGC.
	
	In a more general case, we can consider a flux configuration with spontaneously broken supersymmetry. In particular, for an \textit{elementary saxionic membrane} \cite{lmmv}, that is, a membrane which is only charged under one 3-form in the asymptotic splitting induced by the corresponding asymptotic field direction (e.g. a single D$p$ or NS5-brane wrapping a $(p-2)$ or a 3-cycle respectively in a toroidal orientifold, as the ones in Table \ref{tablaasymp1}), one obtains that $V\sim e^K |W|^2= \m32^2$ (see \cite{fhi, lmmv} for details). Once more, the limit in which the gravitino mass goes to zero implies that the potential also vanishes, and therefore the physical  charge of the corresponding membrane goes to zero and the WGC requires its tension to do the same. The appearance of an infinite tower of states that become light may then be understood in terms of preventing this higher-form symmetry from becoming global once again. Note that, apart from giving a supportive argument for the GDC presented in this paper, this connection between membranes and the scalar potential (both for the supersymmetric and the non-supersymmetric vacua) equally applies to support the ADC.

	\section{Evidence for the GDC in type IIA vacua}
	\label{sec:iia}
	
	In this section we will consider several examples in which the GDC can be analysed in detail.
	In particular, we will see that as the gravitino mass goes to zero there is a tower of KK states
	that becomes light and the corresponding values of $\delta$ can be determined.
	In fact, we will explain that requiring validity of $\cn=1$ supergravity as an effective field theory 
	constrains the range of $\delta$. 
	After showing that the gravitino mass can be written purely in terms of complex structure moduli, we
	will be able to compare the mass scales of KK states and other massive objects, such as strings and membranes,
	depending on the field direction that becomes large.
	We will also describe some runaway dS examples which display a field-dependent gravitino mass 
	directly related to the tension of membranes. 
	
	Let us begin with a succint review of basic results. 
	We will work in the framework of type IIA CY$_3$ orientifolds which give rise to a 4d $\cn=1$ supersymmetric
	theory containing massless chiral multiplets corresponding to the dilaton, plus complex structure and K\"ahler moduli.
	The K\"ahler potential takes the form \cite{Grimm:2004ua}
	\beq
	\label{kpot}
	K= K_K + K_Q, \qquad K_K=-\log (8\cv), \qquad K_Q=4\phi_4 \, . 
	\eeq
	Here $\cv$ is the volume of the internal manifold whereas $\phi_4$ is the 4d dilaton.
	The 10d dilaton $\phi$ and $\phi_4$ are related by $e^{\phi_4}=e^{\phi}/\sqrt{\cv}$.
	
	The mass scale of the various objects that we will consider is set by the Planck mass $\mp$. 
	In particular, the string and KK scales can be written as (see e.g. \cite{fhi})
	\beq
	\label{msmkk}
	\ms = e^{K_Q/4}\ \mp \quad ,\quad  M_{KK}= \frac {\ms}{{\cal V}^{1/6}} = \ms\ e^{K_K/6} \ ,
	\eeq
	Here $M_{KK}$ is the KK scale estimated in terms of the overall volume of the CY$_3$ compactification. 
	
	We now specialize to simple type IIA toroidal orientifolds. We take the internal torus to be factorized, i.e.
	$T^6=(T^2)^3$, and isotropic. Thus, besides the dilaton $S$, there is only one K\"ahler modulus $T$ and one 
	complex structure modulus $U$.
	For this reduced set of moduli the K\"ahler potential $K=K_K+K_Q$ has  
	\beq
	K_K =-3\log(T+\bar T), \qquad   K_Q=- \log (S+\bar S) - 3 \log(U+\bar U) \, .
	\label{kstu}
	\eeq 
	Notice that the internal volume is $\cv=t^3$, where $t={\text{Re}}\, T$. We will also denote
	$s={\text{Re}}\, S$ and $u={\text{Re}}\, U$.
	
	Fluxes are turned on to generate masses for the moduli. The general superpotential induced by R-R, NS-NS and geometric fluxes 
	is found to be
	\beq
	W= e_0 + 3 i e T + 3 c T^2 + i m T^3 + i h_0 S - 3 i h U - 3 a S T - 3 b T U \, .
	\label{wgen}
	\eeq 
	Here we are using the conventions of \cite{cfi}.
	The fluxes $m$, $c$, $e$ and $e_0$ are R-R, while $h_0$ and $h$ are NS-NS. 
	The terms mixing $T$ with $S$ and $U$ arise from geometric fluxes denoted $a$ and $b$.
	Since the theory has $\cn= 1$ supergravity, the F-term scalar potential has the standard expression given in \eqref{eq:VFterm}.
	
	In the following we will need the string and KK scales written in terms of the toroidal moduli.
	Each $T^2$ is generically non-isotropic, having $t=R_x R_y$ and 
	$\tau=R_y/R_x$ as in \cite{fhi}.
	The radii can be expressed in terms of the saxions $t$, $s$ and $u$ using that
	$s=e^{-\phi_4} \tau^{-3/2}$ and $u=e^{-\phi_4}\tau^{1/2}$  \cite{cfi}.
	From \eqref{msmkk} we then obtain
	\beq
	\label{msmp}
	\ms =  \frac{\mp}{(s u^3)^{1/4}},  \qquad 
	M_{KK} =  \frac{\mp}{(s u^3 t^2)^{1/4}}
	\, .
	\eeq
	Since $R_x$ and $R_y$ are not necessarily equal there are actually two separate KK scales defined as 
	$M_{KK}^x=\ms/R_x$ and $M_{KK}^y=\ms/R_y$. It is easy to see that
	\beq
	\frac{M_{KK}^x}{\mp} = \frac{1}{(s  t  u)^{1/2}}\, , \qquad 
	\frac{M_{KK}^y}{\mp} =\frac{1}{u  t^{1/2}}\, .
	\label{KKxy}
	\eeq
	Notice that $M_{KK}^2=M_{KK}^x M_{KK}^y$.  The expressions for $\ms$ and the different KK masses are collected
	in Table \ref{tablaasymp1}.

	\subsection{The gravitino mass and KK towers in type IIA models}
	\label{ss:iiaex}
	
	From the tests of the ADC we know that in supersymmetric AdS vacua
	there are towers of KK states that become light as the cosmological constant $\Lambda$ goes to zero. 
	Since in such vacua $|\Lambda| \sim m^2_{3/2}$, the GDC holds and moreover $\delta=2\lambda$ as discussed before. 
	We further expect the GDC to be valid for non-supersymmetric vacua in analogy with the ADC.
	As we said, we will propose that 
	\beq
	\mkk \simeq  m^\delta _{3/2} \ \ 
	\label{defgrav}
	\eeq
	for some $\delta \le 1$. Moreover, the proposal applies to non-supersymmetric AdS and Minkowski vacua.
	
	In the following we will verify that in a class of type IIA flux vacua the KK states indeed satisfy \eqref{defgrav}.
	We will consider models with universal moduli $S$, $T$ and $U$,  whose K\"ahler potential is given in \eqref{kpot}.
	The superpotential induced by R-R, NS-NS and geometric fluxes has the form \eqref{wgen}. 
	We will discuss AdS and Minkowski vacua, obtained by turning on suitable subsets of fluxes.

	\subsubsection{AdS vacua}
	\label{sss:adsex}
	
	We will present two AdS examples. The first one, dubbed DGKT-CFI, belongs to a class of type IIA models with only NS-NS and 
	R-R fluxes studied originally in \cite{dgkt, cfi}. 
	The second example will include metric fluxes.

	\bigskip
	
	\noindent
	{\bf The DGKT-CFI model}
	\medskip
	\noindent	
	
	The superpotential reads
	\beq
	W=3 i e T + i m T^3 + i h_0 S - 3 i h U \, ,
	\label{wcif}
	\eeq 
	Without loss of generality we take $m >0$.
	It is easy to show that there exists a supersymmetric AdS minimum only if $e < 0$.
	With this choice the moduli are fixed at 
	\beq
	\begin{split}
		{\text{Im}}\, T&=0, \qquad  h_0{\text{Im}}\, S - 3 h {\text{Im}}\, U=0 \, , \\[3mm]
		{\text{Re}}\, T&=t=\sqrt{\frac{5 |e|}{3m}}, \qquad {\text{Re}}\, S=s=-\frac{2 e}{3h_0} t,
		\qquad {\text{Re}}\, U=u=\frac{2 e}{3 h} t
	\end{split}
	\label{vtscfi}
	\eeq
	Since $e<0$, necessarily $h<0$, and $h_0 > 0$.  There are non-zero tadpoles of $C_7$-form, proportional to 
	$mh$ and $mh_0$, that can be cancelled by D$6$-branes and $O6$-planes \cite{cfi}.
	
	At the minimum
	\beq\label{cccfi}
	\frac{\Lambda}{\mpf} \simeq -  \frac {m^{5/2}h_0 |h|^3}{|e|^{9/2}} \simeq -\frac{mh_0}{u^3} \, ,
	\eeq
	where we dropped numerical constants. 
	Substituting the moduli in \eqref{KKxy} gives the KK masses
	\beq
	\frac{M_{KK}^x}{\mpt} \simeq \frac {m^{3/4} (h_0 |h|)^{1/2}}{|e|^{7/4}} 
	\simeq \frac{m^{1/6} h_0^{1/2}}{|h|^{2/3} u^{7/6}}
	\, , \qquad 
	\frac{M_{KK}^y}{\mpt} \simeq \frac {m^{3/4} h}{|e|^{7/4}}
	\simeq \frac{m^{1/6}}{|h|^{1/6} u^{7/6}}\, .
	\label{KKxycfi}
	\eeq
	The overall scale is $M^2_{KK}= M_{KK}^x M_{KK}^y$. 
	The various KK masses have the same dependence on the flux $e$, which is not constrained by tadpole cancellation 
	and can be taken large to ensure the validity of the approximation. Thus we will not distinguish them.
	Since this is a supersymmetric AdS vacuum, \mbox{$\Lambda=-3m^2_{3/2}$} in Planck units. Therefore
	\beq
	\mkk \ \simeq \ |\Lambda|^{7/18}\ \simeq \ m^{7/9}_{3/2} \ .
	\label{wolfe}
	\eeq
	Thus, with the definition in \eqref{defgrav} we have $\delta =7/9$. This is the previously mentioned example that violates
	the (strong) ADC.
	As already remarked, there is some controversy on whether
	this model is a consistent 10d compactification (see e.g. \cite{review} for a review and \cite{Blumenhagen:2019vgj,Font:2019uva,Junghans:2020acz,Buratti:2020kda,Marchesano:2020qvg} for recent assessments). 
	If the model is inconsistent, all the AdS examples 
	in the literature would have $\delta=1$ and no separation between gravitino and KK scales.
	
	We now turn to non-supersymmetric vacua. In \cite{cfi} it was found that in the model with superpotential \eqref{wcif}
	such vacua arise depending on the parameter $\gamma=me$. Up to now we have taken $\gamma < 0$ to have a supersymmetric
	solution. The general situation is much richer though. For starters, there is an AdS  non-supersymmetric stable minimum in
	which the moduli are still given by \eqref{vtscfi} but changing the sign of $e$. Therefore, the KK scales are again given by
	\eqref{KKxycfi}. Concerning $\mg$, it is no longer given by $m^2_{3/2}=|\Lambda|/3$ but 
	it only departs by a numerical prefactor from the supersymmetric value. We then conclude that
	this case satisfies again the GDC, leading to a value of $\delta$ of 7/9.
	
	When $\gamma > 0$ there is a second branch of stable non-supersymmetric AdS vacua in which the vev of the axion
	${\text{Im}}\, T$ is different from zero while the vevs of the saxions have essentially the same form as before.
	Concretely, $hu =-h_0 s$, $2 h_0 s = m t^3$ and $3 m^2 t^2=4\gamma$.
	The KK scales will clearly depend on the unconstrained flux $e$ as in the supersymmetric case and it is easy 
	to show that so does $\mg$.
	Thus, the GDC remains valid for this case with $\delta=7/9$,
	
	It is interesting that all the (perturbatively) stable non-supersymmetric AdS vacua studied above violate the strong
	ADC in that $\lambda< 1/2$.
	It is also remarkable that the GDC holds although without supersymmetry $\Lambda \not= -3m_{3/2}^2$.
	So far the vacua belong to the DGKT-CIF family with $m\not= 0$ and no metric fluxes.
	In the next example these two conditions are relaxed.
	
	\bigskip
	
	\noindent
	{\bf Example with metric fluxes}
	
	\medskip
	\noindent	
	Our discussion here will rely heavily on the analysis of section 4.4 in \cite{cfi}. The superpotential now has the form
	\beq
	W= 3 i e T + 3 c T^2 + i h_0 S - 3aST - 3 i h U -3bTU\, .
	\label{wcif2}
	\eeq 
	Notice that the Roman mass $m$ has been set to zero. The metric fluxes, as well as the NS-NS fluxes are taken to be non-zero
	and to satisfy $h_0=-3 h a/b$.
	
	This model admits supersymmetric and non-supersymmetric AdS minima related by a sign flip in some fluxes.
	In both cases ${\text{Im}}\, T=h_0/3a$ while only a linear combination of ${\text{Im}}\, S$ and ${\text{Im}}\, U$ is
	fixed.
	The saxions are found to be stabilized at
	\beq
	\pm 9 c t^2=-\frac{h_0 e}{a}-\frac{h_0^2 c}{3a^2}, \qquad   s=\frac{2 c}{a} t, \qquad u=\frac{6 c}{b} t \, .
	\label{vevscfi2}
	\eeq
	Observe that the metric fluxes $a$ and $b$ must have the same sign.
	On the other hand, in the expression for $t^2$ choosing either sign gives an extremum solution. 
	A supersymmetric minimum is obtained by taking the plus sign, but we can also choose the minus sign and still get a consistent solution, depending on the fluxes. In this latter case supersymmetry is broken. The minimum is still AdS and it is typically stable \cite{cfi}.
	The grativino mass turns out to be 
	\beq
	m_{3/2}^2=\frac{ab^3}{384 c^2u^3},
	\eeq
	while the KK scales can be obtained by direct substitution of the saxions in \eqref{KKxy}. 
	The important point for us is that they both give the same dependence (modulo a flux-dependent coefficient) on the modulus $t$.
	Specifically
	\beq
	M_{KK}^x \simeq M_{KK}^y \simeq M_{KK} \simeq m_{3/2}=\frac{1}{t^{3/2}}.
	\label{eq:scalesads2}
	\eeq
	Hence we see explicitly that this class of non-supersymmetry vacua would give a value of 1 for $\delta$ in the GDC, provided 
	$t$ can be consistently taken to infinite distance.
	The question is, however, if the large volume limit can be indeed accomplished within the perturbative regime. 
	Since the flux $e$ coming from the 4-form is unconstrained by tadpole cancellation, its absolute value can be taken arbitrarily large
	to guarantee large $t$. In this way the resulting 4d dilaton is small. However, the 10d dilaton
	grows with $t$ and to keep it small requires choosing fluxes appropriately, e.g. letting c to be large enough. However, 
	this seems difficult to realize because the fluxes $a$, $b$ and $c$ are strongly constrained by tadpole cancellation.

	\subsubsection{Minkowski vacua}
	\label{sss:minkex}
	
	The case of Minkowski vacua is compelling because  
	obviously $\Lambda=0$ is fixed and cannot be varied, so the GDC goes beyond the ADC.
	We are not aware of simple explicit Minkowski flux vacua with all moduli fixed but
	there are many known examples of no-scale models with vanishing cosmological constant. 
	Below we will check that the GDC is fulfilled in such models.
	
	Several no-scale examples were presented in \cite{cfi} and others may be designed, both with and without
	metric fluxes. In those models, with universal moduli $S$, $T$, and $U$, both $\mkk$ and $m_{3/2}$ turn out to
	depend on a single no-scale modulus to some power. So there is always a ratio of powers of $\mkk$ and $m_{3/2}$ which only
	depends on fluxes, and indeed one can write a GDC expression like eq.\eqref{defgrav}. There are however 
	no-scale examples with more than one no-scale direction, in which one rather obtains a certain range of values for $\delta$.
	Below we will analyse two models, the first with metric fluxes and the second only with R-R and NS-NS fluxes.
	
	\bigskip
	
	\noindent
	{\bf Minkowski example with metric fluxes}
	
	\medskip
	\noindent	
	The model is defined by the superpotential
	\beq
	W=3 c T^2 + i m T^3 + i h_0 S - 3 a S T \, .
	\label{wiia}
	\eeq 
	This model is of type {\em{NS-4}} in section 4.2 of \cite{cfi}. 
	The scalar potential is positive definite since $W$ does not depend on $U$ and
	$K$ is of no-scale type. There is an extremum with $D_T W=D_SW=0$. 
	Supersymmetry is broken because necessarily $W\not= 0$ so $D_U W\not=0$.
	The generic vevs were found in \cite{cfi}. We take the particular solution 
	\beq
	{\text{Im}}\, T=0, \qquad  {\text{Im}}\, S =0, \qquad
	t=\sqrt{-\frac{h_0 c}{m a}}, \qquad s=-\frac{c}{a} t \, .
	\label{vts}
	\eeq
	The complex structure field $U$ remains undetermined.
	This solution exists for $a c < 0$ and $h_0 m >0$. We choose $a >0$ for concreteness.
	Since the mass matrix has four positive and two zero eigenvalues (from the flat directions) this non-supersymmetric
	Minkowski solution is stable.
	
	The various scales are straightforward to compute. In particular, 
	\beq
	m^2_{3/2} = \frac {a|c|}{32u^3}(\nu+9)\ M^2_{\text{P}}, \qquad
	M^2_{KK}  =  \left(\frac {a^2}{c^2\nu^3}\right)^{\!\!\! 1/4}\ \frac {\mpt}{u^{3/2}} \ .
	\eeq
	where $\nu=h_0 |c|/am$.
	Apparently the GDC is obeyed with $\delta =1/2$.  
	However, we must be careful to also take into account the scales $M_{KK}^x$ and $M_{KK}^y$ which are
	found to be
	\beq
	\frac{M_{KK}^x}{\mp} = \left(\frac{a}{|c|  \nu u}\right)^{\!\!1/2} \, , \qquad 
	\frac{M_{KK}^y}{\mp} = \frac{1}{\nu^{1/4} u}\, .
	\label{KKxy3}
	\eeq
	Then one rather has (in Planck units)
	\beq
	M_{KK}^x\ \simeq m^{1/3}_{3/2} \left( \frac { a^{1/3}}{|c|^{2/3}(\nu+9)^{1/6}}\right) \ ;\ 
	M_{KK}^y\ \simeq m^{2/3}_{3/2} \left( \frac {\nu^{-1/4}}{ [a|c|(\nu +9)]^{1/3}} \right) \ .
	\label{KKxy4}
	\eeq
	As $m_{3/2}\rightarrow 0$, it is   $M_{KK}^y$ which becomes light faster and hence that should be the
	tower which is really relevant for the GDC. So in this type IIA example the result is $\delta=2/3$, rather than 1/2.
	
	Note that one can play somewhat with the flux coefficient in the KK scales, but the fluxes  in $\nu$ and $c$ are 
	strongly constrained by tadpole cancellation.
	Since tadpoles come in the combination $(h_0m-ac)$, none of the fluxes can
	be parametrically large so the coefficients of the GDC are of order one.
	Another interesting result in this example is that the sum of the 4 eigenvalues of the masses of
	$S$ and $T$, i.e. the moduli masses, satisfy 
	\beq
	\sum_{mod}M^2_{mod} \ = \frac {a |c|}{16 u^3}\ 6\nu \ M^2_{\text{P}} \ .
	\eeq
	For large $\nu$ (which means here large $t$), the sum tends to $12m^2_{3/2}$. Thus, the gravitino follows the
	pattern of the moduli, rather than that of the KK states.
	
	\bigskip
	
	\noindent
	{\bf Minkowski example without metric fluxes}
	
	\medskip
	\noindent	
	
	In our last example there are no metric fluxes at all while all R-R fluxes appear.
	Besides, the NS-NS flux $h$ is set to zero so that $W$ is independent of the modulus $U$.
	The superpotential is then  
	\beq
	W= e_0 + 3 i e T +3 c T^2 + i m T^3 + i h_0 S \, .
	\label{wiia2}
	\eeq 
	This example is adapted from the {\em{NS-1}} model in section 4.2 of \cite{cfi}. 
	
	As in the previous model, there is a non-supersymmetric Minkowski solution with
	$D_T W=D_SW=0$ but $D_U W\not=0$. One finds the following vevs for the real and imaginary parts of the moduli
	\beq
	{\text{Im}}\, T=\frac{c}{m}, \qquad  {\text{Im}}\, S =\frac{e_0 m^2+c^3}{h_0 m^2}, \qquad
	h_0 s= m t^3 \, .
	\label{vts2}
	\eeq
	Observe that only a combination of the saxions $t$ and $s$ is fixed, as expected from the absence of mixing between 
	$S$ and $T$ in the superpotential. Necessarily $h_0 m >0$, so that the flux contribution to tadpoles is positive.
	Moreover, this solution exists provided $m\not=0$, $h_0\not=0$, and $\gamma=m e + c^2=0$.
	Regarding the stability of the solution, one can readily show that this extremum must be a minimum because
	the scalar potential is positive definite. Indeed it can be verified that the mass matrix has three positive and three
	zero eigenvalues, the latter due to the flat directions. 
	
	Let us now examine the scales.  For the gravitino mass we find 
	\beq
	\frac{m^2_{3/2}}{\mpt} =  \frac{h_0 m}{32 u^3} \, .
	\label{mgrav2}
	\eeq
	The main novelty now is that the KK scales
	\beq
	\frac{M_{KK}^x}{\mp} = \left(\frac{h_0}{m u}\right)^{\!\!1/2} \frac{1}{t^2} \, , \qquad 
	\frac{M_{KK}^y}{\mp} = \frac{1}{t^{1/2} u}
	\label{KKxy3general}
	\eeq
	turn out to depend not only on $u$ but also on $t$ (or equivalently on $s$)
	because in this case there is an extra modulus which is not fixed by the vacuum condition. 
	
	In order to test the GDC we must limit ourselves to directions in moduli space along
	which the supergravity approximation is reliable. Concretely, we must 
	check that both coupling constants $e^\phi$ and $e^{\phi_4}$ are small enough to remain within the string 
	perturbative regime, and are also compatible with the large volume limit of the internal space. 
	Using  $e^{\phi_4}=e^{\phi}/\sqrt{\cv}$, $e^K=e^{4\phi}/(2\cv)^3$ and $\cv=t^3$ we obtain	
	\beq
	e^{\phi_4} = \left(\frac{h_0}{m}\right)^{\!\!1/4} \frac{1}{(tu)^{3/4}} 
	\, , \qquad
	e^{\phi} = \left(\frac{h_0}{m}\right)^{\!\!1/4} \frac{t^{3/4}}{u^{3/4}} \, .
	\label{dil}
	\eeq
	Hence, naively one could say that in order to have $t \gg 1$, and $e^{\phi},e^{\phi_4} \ll 1$, it is enough to let $u$ tend to infinity 
	while we move in moduli space directions parametrized by the simple power law expression $t \sim u^q$, with $0< q <1$. 
	However, as discussed further in section \ref{ss:limits}, we see that this range is actually restricted to the interval 
	$1/5< q <1/2$. This follows because the fundamental string scale $\ms$ must be larger than both $M_{KK}^x$ and $M_{KK}^y$, and
	we should also ensure that the gravitino scale is below the KK scale ($\delta \le 1$) for the effective field theory to be meaningful.  
	
	Substituting $t \sim u^q$ in the KK masses implies the behavior
	\beq
	\frac{M_{KK}^x}{\mp} \sim \frac{1}{u^{\frac{1}{2}+2q}},   \qquad 
	\frac{M_{KK}^y}{\mp} \sim \frac{1}{u^{1 + \frac{q}{2}}}\, .
	\label{eq:scalesmink2}
	\eeq
	To extract the value of the GDC parameter $\delta$
	we should distinguish two cases here, for depending on whether $q$ is greater or smaller than $1/3$ the lightest mass tower appears 
	to be the one for the $x$ or $y$ direction, respectively. In particular, one can check that for the limiting cases when 
	$q \lesssim 1/2$ ($q \gtrsim 1/5$) the GDC holds for  $\delta \lesssim 1$ ($\delta \gtrsim 11/15$). 
	Also note that when the two directions $x$ and $y$ give identical KK scales (i.e. when $q = 1/3$), the conjecture is fulfilled 
	for $\delta = 7/9$, as in the DGKT-CIF AdS model \eqref{wolfe}.

	\subsection{The gravitino mass and the IIA complex structure sector}
	\label{ss:iics}

	We are interested in the limit in which the gravitino mass goes to zero and we want to identify in which directions in
	moduli space that happens. Examining the examples above one observes that in all cases the gravitino mass 
	can always be written in the form
	\beq
	m_{3/2}^2 \ \sim \frac {(tadpole)}{u^{3} }\ .
	\eeq
	By {\it tadpole} here we mean a bilinear in  integer fluxes contributing to the RR tadpole like e.g. $(h_Im)$ or $(ac)$, etc.  Thus the limit of small gravitino mass in these minima corresponds to the limit with large complex structure.
	
	This is not a particular property of these toroidal vacua but it is always the case for any CY$_3$ orientifold in which
	fluxes contribute to the R-R tadpole. Indeed it is known \cite{Bielleman:2015ina,HIMZ}
	that  for a general type IIA 4d
	${\cal N}=1$, CY$_3$ orientifold the flux scalar potential may be written in the form
	\beq
	V\ =\  V_{\text{3-form}}\ +\ V_\text{loc} \ .
	\eeq
	The first term above comes from the contribution of the different R-R and NS-NS fluxes and is positive definite whereas   $V_\text{loc}$ corresponds to the tensions of
	the localized O$6$ orientifolds (and possibly D$6$-branes). The latter provides for the only negative contribution to the scalar potential
	and imposing tadpole cancellation it may be written as (see e.g.\cite{HIMZ})
	\beq
	V_{\text{loc}}\ =\ \ - \ e^K (\cv) (mh_Ju_J) \ \sim\ -e^{K_Q}(mh_Ju_J),
	\eeq
	e.g. in the case in which only $m,h_J$ contributes to tadpoles. Recall that $\cv$ above denotes the volume of the CY$_3$. We see from here that the local term depends
	only on the complex structure moduli, and not on the K\"ahler moduli. This is expected since the tension of O$6$, D$6$ 
	is proportional to the volume of the wrapped 3-cycles in the CY$_3$. Given this fact, the value of the gravitino mass at the
	minima  will only depend on the complex structure fields.
	The argument is as follows:
	if a minimum is obtained there will be a partial cancellation between 
	the positive terms and the local term. For a minimum  to be reached, the potential at the minimum $V_0$ should scale 
	in the moduli as the piece of the potential coming from the O6-planes.
	But the tension of the orientifold only depends on the complex structure, and hence, at the minimum, 
	$V_0$ may be written in terms of the vevs of complex structure moduli only and one expects (if it is non-zero) that
	$V_0 =\ \sim\ \pm \ e^{K_Q}(mh_Ju_J)$. Consider now the case of ${\cal N}=1$ AdS vacua. In this case
	$V_0=-3m_{3/2}^2$ in Planck units and hence
	\beq
	m_{3/2}^2\ \ \sim\ -\ e^{K_Q}(mh_Ju_J) \ .
	\eeq
	This is in fact true for non-supersymmetric AdS  and Minkowski vacua since the same cancellation between the positive 
	definite terms and the local term has to occur  in order to get a minimum.
	Summarizing, in AdS and Minkowski vacua of type IIA  CY$_3$ orientifolds, the (on-shell) gravitino mass depends only on the
	complex structure and the limit $m_{3/2}\rightarrow 0$ which we are studying corresponds to the large complex structure 
	limit. Nevertheless let us emphasize that, unlike the case of  flux-less  ${\cal N}=2$ vacua, 
	the minimization conditions of the scalar potential may force  the K\"ahler moduli also to be driven to large values.

	\subsection{The gravitino mass and asymptotic moduli limits}	
	\label{ss:limits}	
	
	\subsubsection{General constraints on  $\delta$ from EFT conditions}
	
	According to the GDC we want to study those limits in the moduli space in which the gravitino mass becomes light 
	and the corresponding towers of states. For those limits to make sense we need  the gravitino mass to be smaller  than the
	towers of states involved, so that we can use the formalism of  ${\cal N}=1$ supergravity as an effective field theory (EFT). 
	We also need to have the KK scales of the theory to be lighter than the extended objects in the theory, in particular the 
	fundamental string scale $\ms$. Thus we will require
	\beq
	m_{3/2} \ \lesssim \  M_{KK} \lesssim  \ \ms \ .
	\label{orden}
	\eeq
	Here $M_{KK}$ is the KK scale estimated in terms of the overall volume of the CY$_3$ compactification. 
	Now, from the expressions \eqref{msmkk} one sees that for large moduli the KK scale will always be lighter than $\ms$.
	
	We will also impose that the 10d and 4d dilatons remain smaller than one, so that we stay in a perturbative regime.
	This is
	\beq
	e^\phi\  =\ e^{K_Q/4}\ e^{-K_K/2}\ \lesssim \ 1 \qquad ,\qquad e^{\phi_4}\ =\ \frac {e^\phi}{\sqrt{\cal V}} \ \lesssim \ 1 \ .
	\label{dilaton}
	\eeq
	Here we have used \eqref{kpot} and $e^K=e^{4\phi}/(2{\cal V})^3$. Note that in the large volume regime the 4d theory remains perturbative as long as the 10d theory does.
	Substituting the definition of the overall KK scale, cf. \eqref{msmkk}, one can then obtain a general bound on the exponent $\delta$ 
	which defines the GDC. Indeed, one can write
	\beq
	m_{3/2}^2\ =\ e^K\ |W|^2  \ =\ M_{KK}^6\ e^{-K_Q/2}\ |W|^2 \ >\ M_{KK}^6  \ ,
	\eeq
	since $e^{\phi_4} \lesssim 1$. 
	From this result one concludes that in a type IIA CY$_3$ orientifold at large moduli there is a general
	constraint
	\beq 
	\frac {1}{3}\ <\ \delta \ < 1 \ ,
	\label{lowerbound}
	\eeq
	where the upper limit comes from the EFT condition  $m_{3/2}<M_{KK}$. This general bound is important because it tells us that the
	possible separation of the gravitino scale from the UV scales is bounded.

	\subsubsection{Asymptotic limits of the gravitino mass in toroidal orientifolds}
	\label{sss:asymp}
	
	To gain intuition on the behavior of the towers which become massless in the limit $m_{3/2}\rightarrow 0$,  
	in this section we will consider the class of toroidal type IIA orientifold models studied in \cite{dgkt, cfi} 
	and more recently in \cite{Marchesano:2019hfb, Marchesano:2020uqz}.
	As in section \ref{ss:iiaex}, 
	we will concentrate for simplicity in the isotropic case with the overall moduli being $S,T$ and $U$ with real parts (saxions)  $s,t$ and $u$. 
	Since the gravitino mass depends on the complex structure modulus $u$,  we will take general large moduli limits parametrized as
	\beq
	s\ \sim \ u^r \ \ ,\ \  t\ \sim \ u^q \ .
	\eeq
	General expressions for the masses of KK scales as well other massive objects (particles, strings and membranes) were given in
	\cite{fhi} and are displayed in Table \ref{tablaasymp1}.
	\begin{table}[h!!]\begin{center}
			\renewcommand{\arraystretch}{1.00}
			\begin{tabular}{|c||c|c|c|c|}
				\hline
				Scales & $\ms$ &   $M_{KK}$ & $M_{KK}^x$  & $M_{KK}^y$  \\
				&  $(su^3)^{-1/4}$ & $(su^3t^2)^{-1/4}$  & $(stu)^{-1/2}$ & $(ut^{1/2})^{-1}$  \\
				\hline \hline
				$T_{\text{strings}}$ & D$4(B^0)$ & D$4(B^I)$   & NS$5^a$  &  \\
				&  $s^{-1}$ & $u^{-1}$ & $t^{-1}$ &  \\
				\hline\hline
				$T_{\text{mem}}$ & D$p$ & NS$5^0$ & NS$5^I$ & \\
				& $(su^3t^{(5-p)})^{-1/2}$   &   $(s^{-1}u^3t^3)^{-1/2}$  &   $(st^3u)^{-1/2}$  &  \\
				\hline
			\end{tabular}
			\caption{Masses  and tensions of KK states  and branes in an isotropic $\mathbb{Z}_2\times \mathbb{Z}'_2$ type IIA  orbifold in Planck units.
			}
			\label{tablaasymp1}
		\end{center}
	\end{table}

\begin{figure}
		\centering{}\includegraphics[scale=0.35]{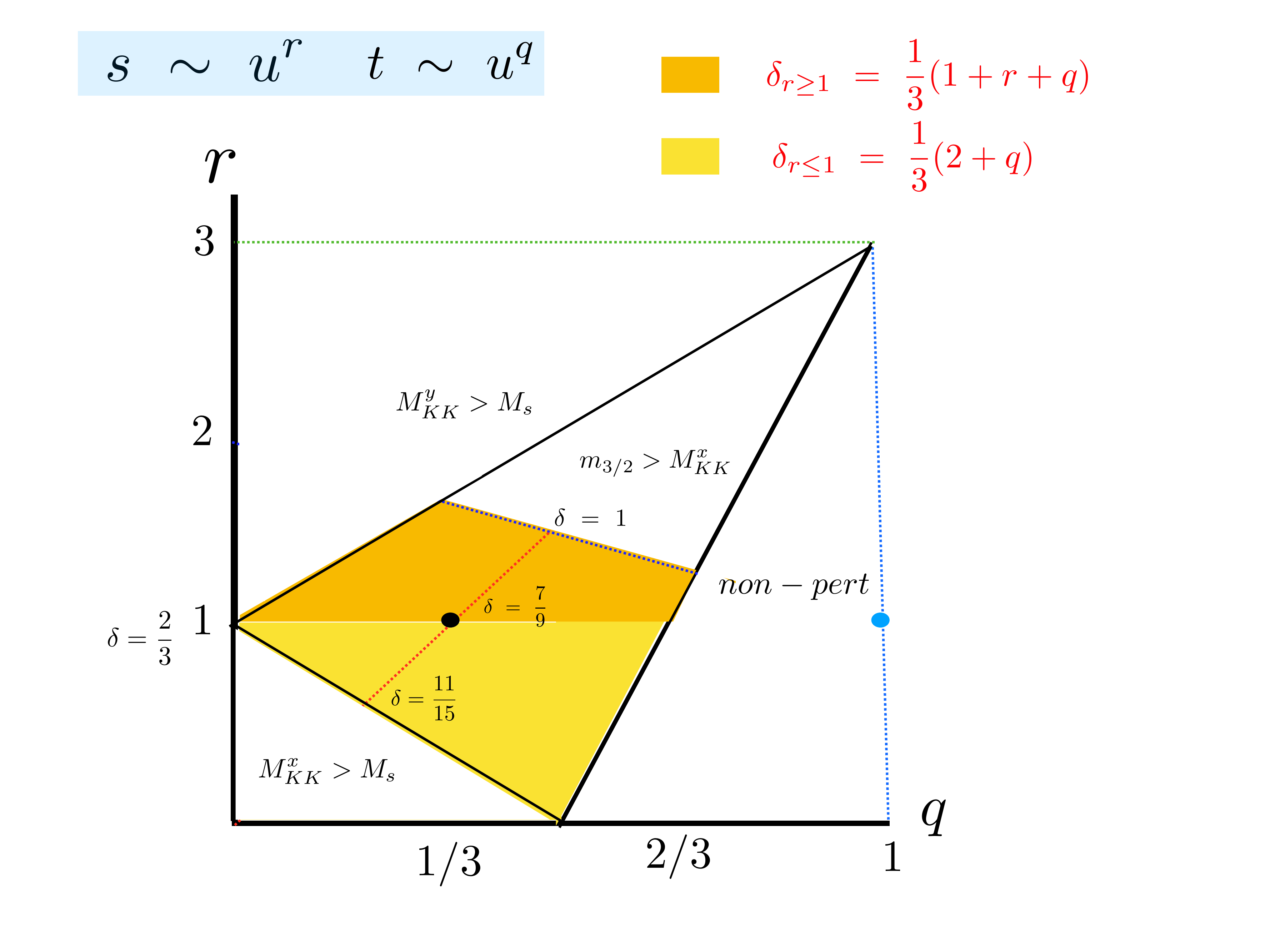} \
		\caption{  Scales in the asymptotic limit  $u\rightarrow \infty$ with
			$s\sim u^r$, $t\sim u^q$. Each point in  the plane $(r,s)$ means one possible field direction. }
		\label{triangulo}
	\end{figure}

	Here we have taken into account that  directions in which $s/u\not=1$ correspond to situations (in a square torus) in which the 
	tori radii $R_x$ and $R_y$ are not equal and hence there are two separate KK scales $M_{KK}^x$ and 
	$M_{KK}^y$ given in \eqref{KKxy}. 
	Let us study first which are the field directions in which there is a well defined perturbative EFT, which requires $e^\phi\lesssim 1$ and
	the constraints  eq.(\ref{orden}). It is easy to check that to stay within the
	perturbative regime requires $r\geq 6q-3$. The general constraints for these $STU$ models are summarized in 
	Figure \ref{triangulo}.  The possible exponents $r,q$ are constrained to be inside the depicted triangle. 
	For  $r>1$ it is $M_{KK}^x$ which provides the lightest KK tower whereas  for $r<1$ it is $M_{KK}^y$. 
	This triangle is purely kinematical and  the moduli directions in any flux compactification of this type 
	should be confined to lie inside in order to get a consistent EFT.

	Let us now make contact with the examples described in section \ref{ss:iiaex}. In general the gravitino mass at the minima scales like $m_{3/2}\sim 1/u^{3/2}$ and imposing that it is lighter than the KK scales cuts the triangle in the middle (line marked $\delta=1$).  
	Then one can write general expressions for the exponent delta, namely
	\beq
	\delta_{r\geq 1}\ =\ \frac {1}{3}(1\ +\ r\ +\ q) \ \ ,\ \ \delta_{r\leq 1}\ =\ \frac {1}{3}(2\ +\ q) \ .
	\label{deltasads}
	\eeq
	Looking at the triangle constraints one sees that any model in this class (either AdS or Minkowski) has $\delta$ in the range
	$2/3\leq \delta \leq 1$.  Here we have defined $\delta$ in terms of the lightest of the two KK scales (either $M_{KK}^x$ or
	$M_{KK}^y$). If one rather considers the $\delta$ corresponding to the subleading tower, lower values like $\delta=1/2$ may be reached.
	Concerning the examples discussed in section \ref{ss:iiaex}, the DGKT-CIF AdS model corresponds to the back dot with $\delta=7/9$.
	The Minkowski examples without metric fluxes in section \ref{sss:minkex} correspond to the dotted red line which has $11/15\leq \delta \leq 1$. 
	The AdS examples with metric fluxes in section \ref{sss:adsex} correspond to the blue dot, which is outside the perturbative regime.

	\subsubsection{Relating the GDC with the SDC}
	\label{ss:SDCandGravitino}

	The goal of this section is to point out the relation between the original SDC \cite{SDC}	and our GDC for Minkowski vacua, in a similar way to the connection between the ADC and the GDC for AdS vacua discussed above. 
	Minkowski vacua are typically characterized by the presence of flat directions associated to the subset of the moduli which are not fixed, as it occurs in the no-scale examples discussed above. Therefore, there is a subspace of the original moduli space which remains flat, i.e. with vanishing potential (at least classically), so that one can freely move within that subspace. The claim is then that as we move in the aforementioned subspace of the moduli space in order to make the gravitino mass tend to zero, we also approach an infinite distance point. The presence of the tower of states which is predicted by the GDC is then guaranteed by the SDC too. Recall the precise form of the SDC (see section \ref{sec:SDC} for more details). If we  now identify the tower predicted by the SDC and the one from the GDC, we can calculate the coefficient $\alpha$ in the SDC from the $\delta$ in the GDC (and viceversa). Writing
	\begin{equation}
	M_{\mathrm{tower}}=c\ m_{3/2}^\delta= c^\prime e^{-\alpha d},
	\end{equation}
	and taking logarithms we obtain
	\begin{equation}
	\log c+ \delta \log (m_{3/2}) = \log c^{\prime} -\alpha d.
	\end{equation}
	Asymptotically, both sides tend to $-\infty$ when the gravitino mass vanishes and the field distance diverges and we can therefore neglect the constant contributions coming from  the first term in each side. Therefore we obtain 
	\begin{equation}
	\label{eq:alphadelta}
	\alpha\, = -\, \delta \, \dfrac{\log (m_{3/2})}{d},
	\end{equation}
	On the other hand we have shown that in general in this class of models $m_{3/2}\sim 1/u^{3/2}$ 
	whereas the proper distance in the one-dimensional subspace spanned by $u$ takes the form
	\begin{equation}
	d=\int \sqrt{2 K_{U \bar{U}}} \, du=\sqrt{\dfrac{3}{2}} \log u + \mathrm{const}.\xrightarrow{u\rightarrow \infty} \sqrt{\dfrac{3}{2}} \log u.
	\label{eq:distance}
	\end{equation}
	Therefore we see that the point at which the gravitino mass vanishes  is indeed at infinite distance, as 
	\beq
	m_{3/2}\ \sim \ e^{-\sqrt{\frac{3}{2}} \,  d} \ \ ,\ \   \alpha\ =\ \sqrt{\frac{3}{2}}\,  \delta \ .
	\label{alphadelta}
	\eeq

	It is important to remark that whereas the prefactor in the exponential of the SDC, $\alpha$, is related to the value of $\delta$, the fact that the gravitino mass is exponentially light with the proper distance is guaranteed as long as $\delta >0$. One could then say that the GDC in Minkowski space  is guaranteed to hold if  the SDC (with a positive $\alpha$) is fulfilled.
	
	Note that the general range we have found for $\delta$, $1/3\leq \delta \leq 1$ then translates to a range for the exponent $\alpha$ of the 
	SDC
	\beq
	\frac {1}{\sqrt{6}} \ \leq \ \alpha \ \leq \sqrt{\frac {3}{2}} \ .
	\eeq
	This inequality is consistent with results of \cite{Grimm1,Corvilain,Gendler:2020dfp,Andriot:2020lea}
	which  studied a different class of asymptotic limits of ${\cal N}=2$ theories in which the 
	towers studied are BPS states becoming massless in the large K\"ahler (rather than complex structure) limit in type IIA.

	\subsubsection{The gravitino mass and tensionless strings and membranes in toroidal type II orientifolds}
	\label{ss:stringsmembranesIIA}

	As we mentioned, the limit $m_{3/2}\rightarrow 0 $ not only drives towers of KK states exponentially massless. Different types of 
	both fundamental and emerging strings, as well as membranes, also become tensionless. The above  class of toroidal orientifold is a good laboratory to 
	explore which extended objects become tensionless when the gravitino gets massless, as well as the rate at which they do.
	
	Let us start with strings. In addition to the fundamental strings, CY$_3$ type IIA orientifolds feature a number of emergent strings which come from branes wrapping cycles in
	the CY$_3$, see e.g. \cite{timo1,timo2,timo4,timoemergent,fhi}.
	In particular there are strings coming from D$4$-branes wrapping (unprojected) 3-cycles in the CY$_3$ and others from NS5-branes wrapping even 4-cycles. 
	In the case of the toroidal orientifold the 4-branes wrap the invariant 3-cycles which we call $B^0$ and $B^I, I=1,2,3$ with tensions given in Table \ref{tablaasymp1}.
	Defining the $\delta$ parameter for string particles as $T_{\text{string}}^{1/2}\sim m_{3/2}^\delta$ it is easy to check that these exponents 
	in this class of isotropic $s,t,u$ orientifolds are given by
	\beq
	\delta_F=\frac {1}{2}(1+\frac {r}{3})\ ,\ 
	\delta_{D4B^I}=\frac {1}{3} \ ,\ \delta_{D4B^0}=\frac {r}{3} \ ,\ \delta_{NS5}=\frac {q}{3} \ ,
	\eeq 
	where the subindex $F$ stands for the fundamental string. 
	From these values it is clear that as $m_{3/2}\rightarrow 0$ both the fundamental strings and the emergent strings from D$4$-branes wrapping the 
	$I=1,2,3$ cycles become tensionless. Concerning the other strings coming form D$4B^0$ and  NS5-branes also typically become tensionless
	except for some field directions with $r=0$ or $q=0$ respectively. We will further comment on the role of tensionless strings at large moduli 
	in section \ref{sec:limits}.
	
	As we already remarked, the $m_{3/2}\rightarrow 0$ limit also drives some membranes tensionless. 
	In particular membranes are obtained from D$p$-branes, $p=2,4,6,8$  wrapping $(p-2)$ cycles in the CY$_3$.
	There are also membranes from the NS5 brane wrapping the $B^0$, $B^I$ 3-cycles. 
	The effect of tensionless 
	membranes is less clear from the point of view of towers of states. One point to remark though is that, as explained in \cite{lmmv},
	tensionless strings appear as boundaries of these membranes and their presence is required by consistency in these compactifications.
	Looking at Table \ref{tablaasymp1} and defining now the $\delta$ exponent as $T_{\text{mem}}^{1/3}=m_{3/2}^\delta$ , one obtains for the exponents
	\beq
	\delta_{Dp}\ =\ \frac {1}{3}(1+\frac {r}{3}+\frac {(5-p)}{3}q)\ ,\ \delta_{NS5^0}=\frac {1}{3}(1-\frac {r}{3}+q)\ ,\ \delta_{NS5^I}=\frac {1}{9}(1+r+3q) \ .
	\eeq
	All these membranes become tensionless as the gravitino mass goes to zero, in agreement with the general arguments that we gave in section \ref{sec:GDC}.

	\subsubsection{The GDC and  dS runaway vacua}
	
	In the previous discussion we have considered AdS and Minkowski vacua, with the gravitino mass defined at the minima or 
	determined by a flat direction in the case of Minkowski vacua.  
	What about dS vacua? We do not have any classical dS vacua other than runaway examples to make a test.
	Still in those vacua one can define a field-dependent gravitino mass in terms of the EFT as $m_{3/2}^2=e^K|W|^2$ and formally explore
	the GDC in those models, studying the connection between the massless gravitino limit and towers of states.
	
	Unlike the case of AdS and Minkowski minima, the field-dependent gravitino mass depends in general in all moduli through the 
	$e^K$ factor, since there is no minima and hence no cancellation of the positive definite terms in the scalar potential 
	with the orientifold contribution to the potential.  The fluxes in general will not contribute to tadpole cancellation 
	in the class of vacua we are considering. Thus the moduli dependence of the gravitino mass will reflect the 
	particular form of the superpotential $W$. Still, it is interesting to consider some simple examples of such 
	runaway flux potentials to compare the gravitino mass scale with the towers of states appearing asymptotically.
	We will take again as a laboratory the class of toroidal type IIA orientifolds considered in previous sections.
	
	We will consider three simple monomial  examples with superpotentials $W_1=e_0$, $W_2=eT$ and $W_3=hU$.
	Taking the asymptotic directions again as $s\sim u^r$, $t=u^q$, one finds for the gravitino masses
	\beq
	m_{3/2}^1\ =\ \frac {1}{u^{(3+r+3q)/2}}\ ,\ 
	m_{3/2}^2\ =\ \frac {1}{u^{(3+r+q)/2}}\ ,\ 
	m_{3/2}^3\ =\ \frac {1}{u^{(1+r+3q)/2}}\ , 
	\eeq
    and correspondingly for the $\delta$ exponents
	\beq
	\delta^1_{r\geq 1}=\frac {1}{3}\frac {(1+r+q)} {(1+r/3+q)}\ ,\ 
	\delta^2_{r\geq 1}=\frac {1}{3}\frac {(1+r+q)} {(1+r/3+q/3)}\ ,\ 
	\delta^3_{r\geq 1}=\frac {1}{3}\frac {(1+r+q)} {(1/3+r/3+q)}\ ,\ 
	\eeq
	\beq
	\delta^1_{r\leq 1}=\frac {1}{3}\frac {(2+q)} {(1+r/3+q)}\ ,\ 
	\delta^2_{r\leq 1}=\frac {1}{3}\frac {(2+q)} {(1+r/3+q/3)}\ ,\ 
	\delta^3_{r\leq 1}=\frac {1}{3}\frac {(2+q)} {(1/3+r/3+q)}\ . 
	\eeq
	Comparing to the case of AdS and Minkowski minima in eq. (\ref{deltasads}) 
	one observes that the $\delta$ exponents are in general smaller,  so that the corresponding 
	KK towers are driven to zero but in a  milder way.
	
	It is interesting to observe that in these monomial examples 
	the implication that membranes become tensionless as $m_{3/2}\rightarrow 0$ is very explicit since one can write
	\beq
	m_{3/2}^1\ = |e_0| \frac {T(D2)}{M^2_{\text{P}}} \ ,\ m_{3/2}^2\ = |e| \frac {T(D4)}{M^2_{\text{P}}} \ ,\ 
	m_{3/2}^3\ = |h| \frac {T(NS5^I)}{M^2_{\text{P}}} \ ,
	\eeq
	in terms of membranes obtained by wrapping appropriate cycles in the internal manifold, as expected.

	For completeness, let us mention here that the models discussed in this section are all compatible with the (refined) de Sitter conjecture \cite{Garg:2018reu, Ooguri:2018wrx} (see also \cite{Dvali:2013eja,Dvali:2014gua,Dvali:2017eba,Dvali:2020etd} for different arguments against the existence of dS in QG). 
	In the present case of runaway examples one only has to check that 
	$|\nabla V| \geq  c V$,
	with $c$ a positive constant of order 1 (in Planck units). The explicit form of the potentials in each case is 
	\beq
	\label{eq:runpot}
	V^1 = 4e^K|e_0|^2,
	V^2 = 4e^K|e|^2 \left[\frac{1}{3} t^2+(\text{Im}\ T)^2\right] ,
	V^3 = 4e^K|h|^2 \left[ \frac{1}{3} u^2+(\text{Im}\ U)^2 \right]  .
	\eeq
	Thus we can compute the norm $|\nabla V|$ of the potential gradient by using the metric on field space given by the kinetic terms of the moduli, i.e.
	\beq
	|\nabla V|=[2 ( \sum_{I,J} K^{I \bar{J}} \nabla_I V \ \nabla_{\bar{J}} V)]^{1/2} ,
	\eeq
	where $K^{I \bar{J}}$ is the inverse K\"ahler metric in moduli space. In the asymptotic regime, with $u\rightarrow \infty$, $s\sim u^r$, $t\sim u^q$, where also the contribution to the scalar potential from their axionic partners can be neglected, one can check this to be proportional to $V$, with a flux independent constant $c$ equal to $(\sqrt{14}, \sqrt{26/3}, \sqrt{26/3})$ for the examples 1), 2) and 3), respectively. Note that this basically amounts to restrict ourselves to the ``vacuum" condition, which fixes $\text{Im}\ T=0$ or $\text{Im}\ U=0$. However, it is also interesting to mention that even though the dS conjecture seems to be fulfilled asymptotically, this also holds even when considering the axions outside the minima, since $|\nabla V|$ will be greater than that of the minimum, and thus $|\nabla V|/V>c$ is still satisfied. Hence, the constants $c^i$ computed above really give a lower bound for the constant appearing in the dS conjecture. Also note that these bounds agree with some general no-go theorems in type IIA geometric compactifications \cite{PKTT}, which state that in such compactifications in string theory with fluxes and orientifolds any $V>0$ satisfies $|\nabla V|/V \geq \sqrt{54/13}$.

	Let us  end this section by making  a couple of comments regarding the Higuchi bound \cite{Higuchi:1986py,lp}. This is a consistency condition that must be satisfied when considering dS backgrounds. Unitarity of massive representations of higher spin particles imposes a bound, which implies that the lightest KK replica of the graviton must obey
	\beq
	\label{eq:higuchi}
	\mkkt \, \geq \, 2H^2\, =\, 2\frac {V_0}{\mpt} \, .
	\eeq
	On the other hand  if  the GDC is  fulfilled by the KK compactification scale, then $\mkk \simeq \m32^\delta$ and one would have
	\beq
	\label{eq:higuchi2}
	\m32^{2\delta} \, \geq \, 2\dfrac{V_0}{\mpt} \ .
	\eeq
	This means that as the gravitino mass goes to zero, $V_0\rightarrow 0$ also, which is consistent with the expectation that one should recover supersymmetry when the gravitino approaches the massless limit, and hence Minkowski. Note also that then a {\it dS distance conjecture} applies since,  when the limit $V_0\rightarrow 0$ is
	taken, it comes along with a tower of massless states, which is the one of the GDC\footnote{We thank D. L\"ust and C. Vafa for discussions on this issue.}.
	
	Note that  in the previous runaway examples the Higuchi bound is obeyed although, being non-stable vacua,  it is not obvious that the conditions of unitarity on which it is based should apply. However, if one insists in any case in the Higuchi bound to be satisfied in these examples, one can see that our previous requirement of $\delta \leq 1$ for the EFT to be under control is actually in agreement with this unitarity bound in dS.  Thus, if one has $V_0\rightarrow 0$, eqs. \eqref{eq:higuchi} and \eqref{eq:higuchi2} above, combined with the fact that our runaway potentials \eqref{eq:runpot} satisfy $\m32^2 \simeq V_0$, require  $\delta \, < \, 1$. Thus, in this sense, the Higuchi bound is compatible with the GDC and moreover imposes no further restrictions to its most relevant parameter $\delta$.

	\section{The gravitino mass in the limits of moduli space}
	\label{sec:limits}

	The analysis presented up to this point in order to show evidence in favour of the GDC has focused mostly on type II compactifications to 4d in some specific toroidal orientifold models. We have discussed examples of vacua in the large volume approximation in IIA (large complex structure in IIB) and weak coupling limit so as to remain within the regime of applicability of the effective supergravity action. Hence, the following question arises quite naturally: can we be more general in our assertions to include a broader class of vacua obtained through generic CY$_3$ orientifold compactifications and still be able to give explicit values of the relevant parameters appearing in the GDC? If this is to be case, we could then perhaps give some model-independent constraints on the parameter $\delta$ in the GDC, or extend our results to other infinite distance points in moduli space apart from the commonly explored ones. 

	In this section we will try to address these questions by using the machinery of limiting Mixed Hodge Structures (MHS). In the next subsection we will present the main results and formulae to be used in the upcoming ones. This formalism will give us some tools to treat (some of) our type II asymptotic vacua examples of section \ref{sec:iia} in a unified way and extract the general features appearing in those simple but instructive models. However, before jumping straight into the discussion that is to follow, a cautionary note is in order. The framework of asymptotic Hodge theory is well adapted to discuss independently the complex structure \cite{Grimm2} and (the mirror) K\"ahler structure sectors \cite{Corvilain} of type II string compactifications. However, we have seen in our examples that the on-shell gravitino mass does not seem to depend on the K\"ahler moduli in type IIA (complex structure in type IIB). Still, in order for the aforementioned mass to approach zero while having the effective description under control, we saw that we needed to sample directions in moduli space with both sectors taken to the boundary.

	The focus will be therefore to study those models that can be lifted to flux compactifications in F-theory on CY$_4$, in the spirit of \cite{Grimm:2019ixq}. This will be briefly reviewed in the next subsection. The reason for concentrating on F-theory compactifications is that it let us study examples in type IIA supergravity (after taking the suitable limits and making proper identifications through mirror symmetry) where both the dilaton and K\"ahler moduli are taken to infinity.  Moreover, we will restrict ourselves to codimension-two loci in the boundary of moduli space. The reason for this is twofold. First, these cases were extensively analysed in \cite{Grimm:2019ixq}, and we can then apply  their results directly to our setting. Second, by relaxing some constraints on the asymptotic scalar potential in the dual type IIA theory, they showed that the only possible AdS vacua were precisely the infinite family studied in \cite{dgkt,cfi}, which we have discussed earlier in section \ref{ss:iiaex}.

	\subsection{Rudiments of Asymptotic Hodge Theory}
	\label{ss:MHS}
	
	The \emph{raison d'etre} of this section is to provide the reader with the main results and applications of \emph{Asymptotic Hodge Theory} to string compactifications and the {\it Swampland Program} \cite{Grimm1, Grimm2, Corvilain,Grimm:2019wtx,Grimm:2019bey,Grimm:2019ixq,Grimm:2020cda,Grimm:2021ikg,Bastian:2020egp,Grimm:2020ouv}. This will serve us to set also the notation that will be used in the discussion that is to follow.
	
\begin{figure}[htb]
		
		\begin{center}	
			
			\includegraphics[scale=0.45]{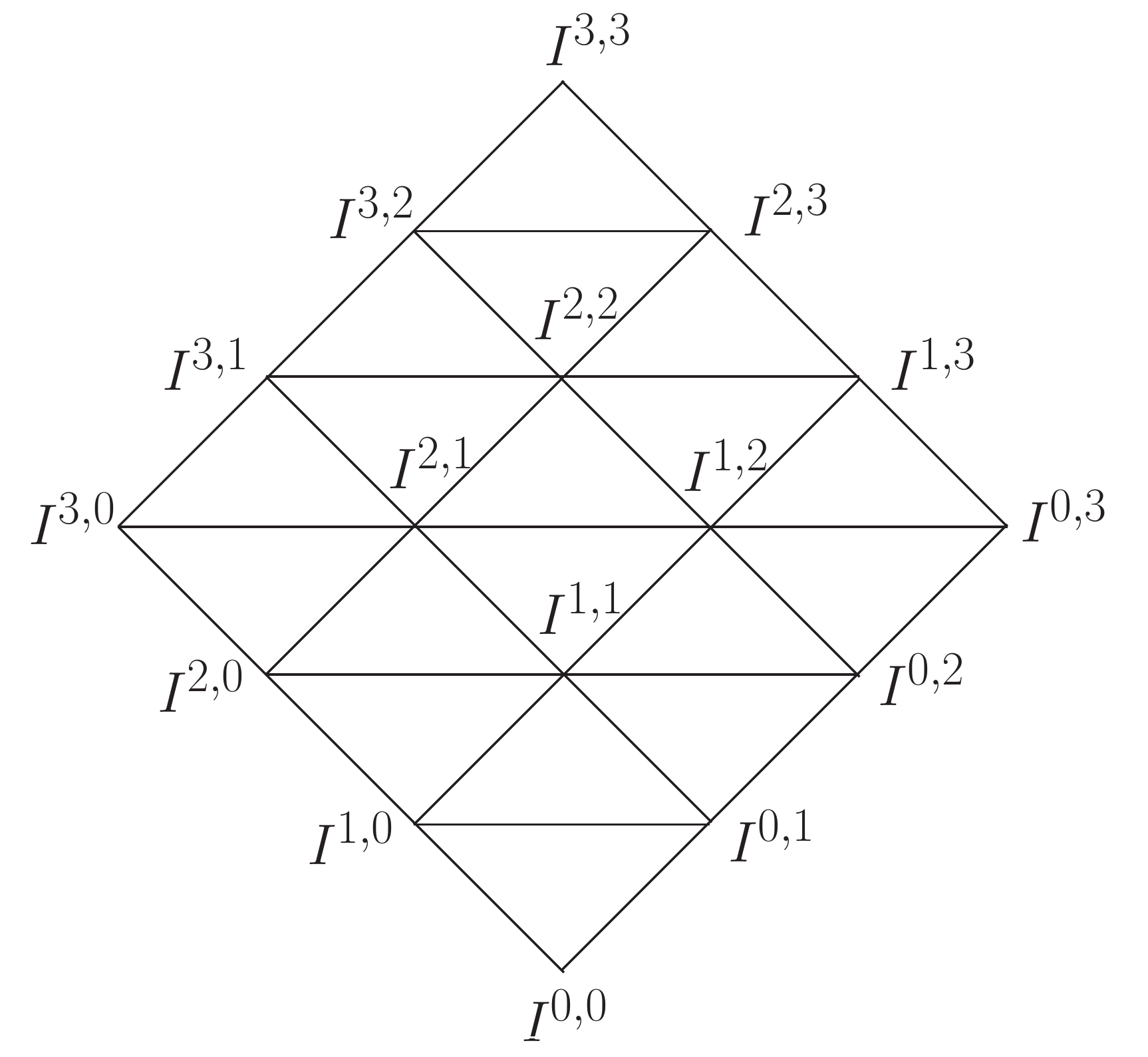}
			
		\end{center}
		
		\caption{Limiting Hodge diamond for the middle cohomology $H^3 (\text{CY}_3, \mathbb{C})$ of an unspecified CY$_3$.}\label{fig:HDIpq}
		
	\end{figure}

	One of the main observations about the moduli space of CY compactifications, denoted here as $\mathcal{M}_{mod}$, was that it is neither smooth nor compact. In fact, it was shown \cite{quasiprojective} that there are special loci where the compactification manifold becomes singular. Those points were seen to lay at both finite (e.g. the conifold \cite{Vafa:1995ta,Strominger:1995cz}) and infinite distance (measured with the metric defined in moduli space). The latter are the main goal of some of the most recently studied Swampland Conjectures, as the SDC (see section \ref{sec:SDC}), and thus one of the main focus in this work. Locally, each singular locus can be described as the intersection of several divisors. One can thus construct a set of local (complex) coordinates in moduli space, $T^a=t^a+ i b^a$, with $a=1,\ldots, \text{dim}_{\mathbb{C}} (\mathcal{M}_{mod})$,  such that this intersection is characterized by some subset of them having infinite real part, i.e. 
	\beq
	T^j=t^j + i b^j , ~~~~~~T^j\rightarrow \infty\, 
	\eeq
	where the index $j=1,\ldots , \hn$ (with $\hn \leq \text{dim}_{\mathbb{C}} (\mathcal{M}_{mod})$) denotes the subset of the moduli which are taken to infinity. Notice that the real (imaginary) part $t^j$ ($b^j$) corresponds here to the saxion (axion) of the associated 4d field. For definiteness we will refer henceforth to the complex structure moduli space, $\mathcal{M}_{cs}$, of a general CY $d$-fold ($\text{CY}_d$), but one should note that similar arguments let one analyse singular limits in other sectors, such as the K\"ahler structure sector of its mirror compactification \cite{Corvilain}. Hence, all the relevant information is encoded in the holomorphic $(d,0)$-form $\Omega(T^a)$, which depends on the complex structure local coordinates. This form belongs to the middle cohomology $H^d (\text{CY}_d, \mathbb{C})$ of the internal (complex) manifold, such that for fixed complex structure, it can be decomposed into a direct sum as follows
	\beq
	H^d (\text{CY}_d, \mathbb{C})= H^{d,0} \oplus \ldots \oplus H^{0,d}\, .
	\label{hodge}
	\eeq
	As a consequence of being at the singularity, though, the above splitting breaks down. However, one can still extract some refined mathematical structure captured by the so-called \emph{Deligne splitting} \cite{Deligne}, $H^d (\text{CY}_d, \mathbb{C})=\bigoplus_{p,q=0}^{d}  I^{p,q}$. Each of these subspaces $I^{(p,q)}$ has dimension $i^{p,q}=dim_{\mathbb{C}}(I^{p,q})$, which are related to the usual Hodge numbers $h^{p,q}$ by the following expression
	\beq
	\sum_{q} i^{p,q}=h^{p,d-p}.
	\eeq
	This may be depicted by the \emph{limiting Hodge diamond} displayed in Figure \ref{fig:HDIpq}.
	
	To build this splitting we need some local information near the singular locus. The central element in this respect is the \emph{monodromy matrix}, which captures the behaviour of the holomorphic $(d,0)$-form when one encircles the singularity, i.e. under the axionic shift $T^j \rightarrow T^j+i$. It is convenient at this point to look at the period vector, usually denoted as $\mathbf{\Pi},$ that comprises the coefficients of $\Omega$ when expanded on an integral basis $\mu_I$, $I=1,\ldots, 2 h^{d-1,1}+2$, of $H^d (\text{CY}_d, \mathbb{Z})$, that is 
	\beq
	\Omega=\Pi^I \mu_I.
	\eeq
	The periods are not uniquely determined and undergo some transformation $\mathbf{\Pi} (T^j+i)= R_j \ \mathbf{\Pi}(T^j)$ via the aforementioned shift, which is encapsulated by the monodromy matrix $R_j$. However, the interesting object is not the monodromy matrix itself but its logarithm, $N_j= \text{log}(R_j)$, which can be seen to be nilpotent \footnote{Strictly speaking one has to first extract the unipotent part of the monodromy matrix, which can be reached after a change of basis \cite{Grimm1}.}, i.e. that there exists some $m_j \in \mathbb{Z}$ such that $N_j^{m_j} \neq 0$ and $N_j^{m_j+1} = 0$. What is most striking about this is the fact that the nilpotent matrix captures a great deal of information about the singularity. Hence, by means of the \emph{Nilpotent orbit theorem} \cite{nilpotent} one can write the above period vector as
	\beq
	\mathbf{\Pi}(T^j)= e^{-i T^j N_j} \mathbf{a_0}+\mathcal{O}\left(e^{-2 \pi T^j} \right)
	\label{eq:nilpotent}
	\eeq
	where both $\mathbf{\Pi}$ and $\mathbf{a_0}$ may additionally depend on the non-divergent moduli. Several comments are in order. First, it should be noted that the nilpotent orbit lets us extract the singular part of the period vector, which explicitly depends on the nilpotent matrix $N_j$ (notice that under the shift $T^j \rightarrow T^j+i$, the above formula states that $\mathbf{\Pi}$ transforms by the monodromy matrix $R_j$). The order $\mathcal{O}\left(e^{-2 \pi T^j}\right)$ terms are subleading in the asymptotic expansion, and thus can be neglected when approaching the singular locus $T^j \to \infty$. Finally, one should note that by the nilpotency property of $N_j$, the exponential in \eqref{eq:nilpotent} only contains a finite number of terms, leading to a polynomial dependence on the divergent moduli. 
	
	Moreover, the nilpotent matrix can be used to give a classification of the allowed singularities within the moduli space under consideration \cite{Grimm1, Grimm2,Grimm:2019bey}. This is done by studying the degree of the above polynomial, i.e. which $d_j \in \mathbb{Z}_{\geq 0}$ fulfills the condition 
	\beq
	\label{eq:defd_j}
	(N_j)^{d_j} \mathbf{a_0} \neq 0, ~~~ \text{and} ~~ (N_a)^{d_j+1} \mathbf{a_0} = 0. 
	\eeq

	The latter is usually indicated by referring to the corresponding singularity as one of type I for $d_j=0$, type II for $d_j=1$, etc. For CY$_3$ and CY$_4$ this classification is further refined adding some subindices to the singularity type (see e.g. \cite{Grimm2,Grimm:2019ixq}).
	
	Interestingly, when several moduli are sent to infinity (i.e. for codimension greater than $1$ singular loci), path dependence becomes important, and one has to specify the order in which one approaches the boundary. Hence, one obtains a different limiting Hodge diamond at each step of the (accumulated) singularity. This provides us with an enhancement chain \cite{Grimm2} characterized again by the nilpotent matrix $N_{(j)}=N_1+\ldots+N_j$, as shown below \cite{Grimm2}

	\beq
	\xrightarrow{t^1 \to \infty} \text{Type} \ X_{(1)} \xrightarrow{t^2 \to \infty} \text{Type} \ X_{(2)} \xrightarrow{t^3 \to \infty} \ldots \xrightarrow{t^{\hn} \to \infty} \text{Type} \ X_{(\hn)}
	\label{chain}
	\eeq
	The nilpotent orbit leaves an expression for the period vector which is still difficult to deal with. In order to simplify further one can introduce a \emph{growth sector}, which amounts to divide the local patch around the singularity into disjoint sectors by introducing a specific ordering, namely
	
	\beq
	\label{growthsector}
	\mathcal{R}_{1,2, \ldots \hat{n}}=\left\{ T^j=t^j+ i b^j \mid \frac{t^{1}}{t^{2}}>\gamma, \frac{t^{2}}{t^{3}}>\gamma, \ldots, \frac{t^{\hn -1}}{t^{\hn}}>\gamma, t^{\hn}>\gamma, b^{j}<\delta\right\}_{\gamma \gg 1, \delta>1}.
	\eeq
	The analysis that follows will be valid for any path belonging to the specified growth sector. As long as one stays inside it, the results from the \emph{sl(2)-orbit} can be applied as in \cite{Grimm2}. 
	
\begin{figure}[htb]
		
		\begin{center}
			
			\includegraphics[scale=0.40]{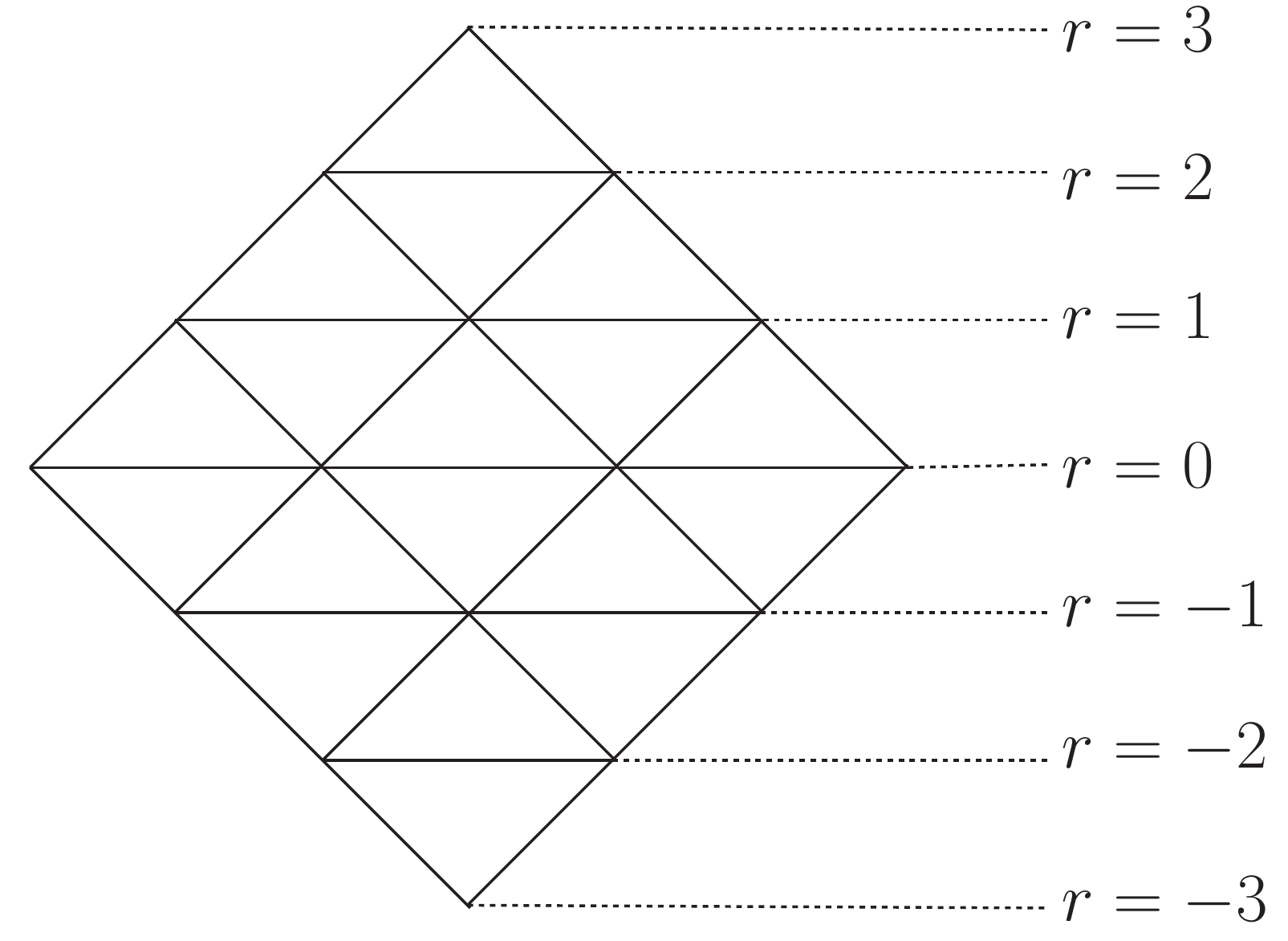}
			
		\end{center}
		
		\caption{Limiting Hodge diamond displaying the $r_j$ levels in which generic $d$-forms live in the singularity. For definiteness, we have particularized here to the case $d=3$.}\label{fig:HDrs}
		
	\end{figure}

	Recall that the nilpotent orbit approximation discarded exponential corrections to the period vector $\mathbf{\Pi} (T^a)$. The sl($2$)-orbit further neglects subleading polynomial corrections in $t^j/t^{j+1}$, which are obviously suppressed within the chosen growth sector. Moreover, one can construct a particular splitting of the real cohomology from the $\mathbb{R}$-split Deligne splitting (see \cite{Grimm2})
	\begin{equation}
	H^d (\text{CY}_d, \mathbb{R})= \bigoplus_{\mathbf{r}=(r_1, \ldots ,r_a) } A_{\mathbf{r}} ,
	\label{eq:realsplitting}
	\end{equation}
	where $-d \leq r_j \leq d$. Note that to make this splitting one has to restrict oneself to the real cohomology as shown in Figure \ref{fig:HDrs}.
	
	Finally, another important concept is that of the \emph{Hodge norm}. This is inherited from the inner product between forms in the internal manifold
	\begin{equation}
	|| \alpha ||^2 = \left < \alpha , \alpha \right> = \left( \int_{X_d} \alpha \wedge *  \alpha \right) ,
	\label{hodgenorm}
	\end{equation}
	where $X_d$ denotes the compactification manifold and $*$ the Hodge-star operator associated to its metric. We will have special interest in the Hodge norm of forms belonging to the middle cohomology $ H^d (\text{CY}_d, \mathbb{R})$, the reason being that the holomorphic form $\Omega(T^a)$ (and also the 4-form fluxes that we will consider later in next section) enter in this category. Moreover, the Hodge star operator acts within this space as a multiplicative factor \cite{Grimm2}. The point here is that this norm also gets simplified if we use the sl($2$)-orbit approximation within a given growth sector. Hence, if $\alpha \in A_{\mathbf{r}}$, then the Hodge norm reduces simply to
	\beq
	|| \alpha ||^2 \sim \left(\frac{t^1}{t^2}\right)^{r_1}\left(\frac{t^2}{t^3}\right)^{r_2}\ldots (t^{\hn})^{r_{\hn}} \ ,
	\eeq
	where again not only exponential but also polynomial corrections in the divergent moduli have been neglected, and $\sim$ indicates that we are not paying attention to finite prefactors. It is worth mentioning too that the $A_{\mathbf{r}}$ subspaces are `orthogonal' in the sense that for $\alpha \in A_{\mathbf{r}}, \ \beta \in A_{\mathbf{r'}}$ we have $\left < \alpha , \beta \right>=0$ unless $\mathbf{r}+\mathbf{r'}= 0$. This turns out to be a key property to approximate quantities asymptotically, such as K\"ahler potentials that take the form  $K=- \text{log} \left < \mathbf{\Pi} , \bar{\mathbf{\Pi}} \right> \simeq - \text{log} \left( (t^1)^{d_1} (t^2)^{d_2-d_1}\ldots (t^{\hn})^{d_{\hn}-d_{\hn-1}}\right)$.
	
	\subsection{Application to F-theory flux compactifications}
	\label{ss:fthy}
	
	Here we want to briefly recall how flux compactifications of 11d supergravity (low energy limit of M-theory) on (smooth) CY$_4$ relate to some type IIB/A orientifold compactifications in special limits, following closely the exposition in \cite{Grimm:2019ixq} (see also e.g.\cite{Becker:1996gj,Dasgupta:1999ss} for more details). Readers familiar with this material may jump to section \ref{ss:vacua2mod}. 
	
	Consider M-theory compactified on a CY$_4$ dubbed $X_4$, leading to 3d $\mathcal{N}=2$ supergravity. We allow for some $G_4$ flux along some internal four cycles on the CY$_4$. This leads to a scalar potential of the form
	\beq
	V_\text{M} = \frac{1}{V_4^3} \left( \int_{X_4} G_4\wedge *  G_4 -\int_{X_4} G_4\wedge G_4 \right) ,
	\eeq
	with $V_4$ the volume of $X_4$. Additionally, there is a consistency relation linking the flux $G_4$ and the curvature of the internal manifold \cite{Haack:2001jz}.
	We will be mostly concerned with vacua determined by the complex structure moduli $z^J$, $J=1,\ldots,h^{3,1}(X_4)$ of the internal manifold. Hence, we further restrict to 4-form fluxes satisfying the primitivity condition $J \wedge G_4=0$ (i.e. $G_4 \in H_{\text{p}}^4 (X_4, \mathbb{R})$). As a consequence, $V_\text{M}$ will depend only on the complex structure moduli and the overall volume factor, rendering the potential positive definite due to its no-scale condition. In supergravity language, the potential can be then rewritten in terms of a K\"ahler potential $K$, and a superpotential $W$ as follows 
	\beq
	V_\text{M}= e^K K^{I \bar{J}} D_I W  \overline{\left(D_{J} W\right)} ,
	\label{Mpot}
	\eeq
	where, as usual, $D_IW=\partial_IW+K_IW$ and $K_{I\bar{J}}$ is the K\"ahler metric in the complex structure moduli space $\mathcal{M}_{cs}$ of $X_4$. For completeness, we recall here the usual expressions both for the K\"ahler potential,
	\beq
	K = K^{cs}(z,\bar{z})-3 \text { log }(V_4), ~~~~~  K^{cs}(z,\bar{z})=-\text { log } \left(\int_{X_4} \Omega(z)\wedge \bar{\Omega} (\bar{z}) \right ),
	\eeq
	and the  Gukov-Vafa-Witten superpotential \cite{Gukov:1999ya}
	\beq
	W =\int_{X_4} G_4 \wedge \Omega(z) .
	\eeq
	We proceed now sketching the chain of dualities that lets us relate this potential to an analogous one in type II orientifold compactifications. One can see that the above potential lifts almost directly to an F-theory scalar potential if $X_4$ is taken to be an elliptically fibered CY$_4$ (with base denoted by $B_3$) by shrinking the volume of the torus fibre to zero. Hence one obtains the following scalar potential in F-theory
	\beq
	V_\text{F} = \frac{1}{V_b^3} \left( \int_{X_4} G_4 \wedge * G_4  - \int_{X_4} G_4\wedge G_4  \right) , 
	\eeq
	where $V_b$ denotes the volume in the 10d Einstein frame of a type IIB compactification over $B_3$. 
	
	Finally, one can take Sen's weak coupling limit \cite{Sen:1996vd}, to describe a type IIB CY$_3$ orientifold compactification over the threefold base $B_3$. If one further goes to the string frame and makes use of mirror symmetry, the following type IIA orientifold (in the mirror $\tilde{X}_3$ CY$_3$) scalar potential can be obtained \cite{Grimm:2019ixq}
	\beq
	V_{\text{IIA}} = \frac{1}{4s^3 \left|\Omega^A\right|^4} \left( \frac{s}{t^3}\left|\Omega^A\right|^2 \int_{\tilde{X}_3} H_3\wedge * H_3 + \frac{1}{st^3} \sum_{p ~\text{even}} \int_{\tilde{X}_3} F_p\wedge * F_p - \int_{O6/D6} F_0 H_3 + \ldots\right). 
	\eeq
	Note that, as customary, we have  labelled the saxion corresponding to the dilatonic chiral multiplet as $s$, and the K\"ahler modulus $t$. From the above formula it is obvious that it correctly reproduces the contributions to the scalar potential arising from the NS-NS $H_3$ form flux and R-R $F_p$ form fluxes considered in e.g. \cite{cfi}, while the dots denote some extra contributions coming from geometric and non-geometric fluxes which will be taken into account later on. 
	
	Before proceeding we should make a couple of important observations. First of all, from duality with M-theory we recover also the piece coming from the tensions of the localised sources, i.e. O$6$ planes and D$6$ branes. This last term will be the only negative contribution. Second, since this potential is dual to the one introduced in \eqref{Mpot}, which as we said was manifestly positive-definite, there must exist some correlation between the local term and the fluxes. However, as was argued in \cite{Grimm:2019ixq}, one can relax this correlation between the flux parameters so that more general potentials in type IIA, with the same dependence on the two moduli that will be sent to the limit (i.e. $s,t$), can be described. Thus non-positive definite potentials can also be analysed, as required to recast our AdS vacua within this framework. However, in order to do that we will have to take $s\sim u$ in our examples (which is indeed compatible with the vacuum conditions, see section \ref{ss:iiaex}), so as to have an effective two-moduli problem. 
	
    \subsubsection*{Two-moduli limits in F-theory}

	Henceforth we will restrict ourselves to the two-moduli case, as anticipated before. We will also make use of the main tools of MHS reviewed in section \ref{ss:MHS}. Along the way we will see that the form that the type IIA potential adopts in the asymptotic limits can be readily identified with the ones we have used in this paper, provided we set $s\sim u$ and a new ``free" parameter $\eta$ that will be introduced later on in this subsection equal to 3 \cite{Grimm:2019ixq}. 
	
	Hence consider the complex moduli space of a CY$_4$ with $h^{3,1}=2$ and send both coordinates to a singular limit (see \cite{Grimm:2020ouv} for more details), which can be locally parametrized as
	\beq
	S= s + i b^1 , ~~~ T=t + i b^2,
	\eeq
	such that the singularity is reached if $s,t \rightarrow \infty$. Also we fix a growth sector (c.f. \eqref{growthsector})
	\begin{equation}
	\mathcal{R}_{s,t}=\left\{ (s,t) \mid \frac{s}{t}>\gamma,  b^{i}<\delta\right\}_{\gamma \gg 1, \delta>1}.
	\label{growthsector2}
	\end{equation}
	
	Note that we could exchange the coordinates so as to consider the alternative growth sector. However, we will be more interested in this particular ordering for reasons that will become clear in next subsection. It is important to keep in mind that we are taking $\gamma$ to be very large so as to be able to use the \emph{strict asymptotic approximation} (see discussion above \eqref{eq:realsplitting}), in which one discards exponential and subleading  polynomial (in $s/t$) corrections, which in this context can be interpreted as perturbative and/or non-perturbative $\alpha'$ corrections.
	
	Recall that we could classify the singularity type as well as the allowed enhancement by studying the corresponding monodromy matrix. For a complete list in this two-moduli limit case we refer the reader to tables 5.1, 5.2 in \cite{Grimm:2019ixq} (and the original work \cite{Kerr}). Regarding the $\mathbb{R}$-split of \eqref{eq:realsplitting}, one is interested in this case in the one corresponding to the primitive part of the middle cohomology, $H_p^4 (X_4, \mathbb{R})$, of the CY$_4$, as the $G_4$ flux belongs to this class
	
	\begin{equation}
	H_p^4 (X_4, \mathbb{R})= \bigoplus_{\mathbf{r}=(r_1,r_2) \in \Gamma} A_{r_1 r_2},
	\label{eq:4realsplitting}
	\end{equation}
	where the set $\Gamma$ of possible $(r_1,r_2)$, with $-4 \leq r_1,r_2 \leq 4$, depends on all this data.
	
	A crucial concept for us will be that of \emph{unbounded asymptotically massless} (UAM) fluxes \cite{Grimm:2019ixq}. These denote those components in $A_{r_1 r_2}$ above which have the following properties
	\begin{itemize}
		\item $\left < A_{r_1 r_2},A_{r_1 r_2} \right >=0$ ,
		\item $\left < A_{r_1 r_2},G_4^{\text{rest}} \right >=0$ ,
		\item $\left\| A_{r_1 r_2} \right\| \rightarrow 0 ~~ \text{on every path with} ~ s,t \rightarrow \infty$\, ,
	\end{itemize}
	
	where $G_4^{\text{rest}}$ above refers to the rest of the components of the 4-form flux which are not unbounded. The significance of these UAM fluxes is that they do not contribute to the tadpole cancellation condition and hence, at least from this perspective, are not bounded, and also their contribution to the potential asymptotically vanishes (they produce a mild backreaction).

	\subsection{Type II vacua in codimension-two boundaries}
	\label{ss:vacua2mod}
	
	Now it is time to apply these results to gain some information about the asymptotic scalar potential in type II orientifold compactifications. Recall that we are particularizing to the two-moduli case here. Thus, we have two scalar fields, which in type IIA are taken to be the dilaton $s$ and the (universal) K\"ahler structure modulus $t$, both becoming large as we approach the boundary. We also need to specify a growth sector in order to apply the strict asymptotic approximation, which in our case will be $\mathcal{R}_{s,t}$, whose paths are characterized by $s$ growing faster than $t$. Note that the motivation for this becomes clear once one takes $s\sim u$, as happened in our AdS examples, and asks for the supergravity theory to stay within the string perturbative regime. Indeed by substituting in \eqref{dilaton}, we see that the 10d dilaton vev goes like $\ e^\phi \mp \sim \mp \frac{t^{3/2}}{s}$, justifying our choice (see discussion in sect. \ref{sss:asymp}).
	
	We start then from the generic asymptotic potential for the codimension-two limit
	\beq
	V_\text{{IIA}} = \frac{1}{s^\eta} \left( \sum_{(r_1, r_2)\in \Gamma}V_{r_1 r_2}  \ s^{r_1}t^{r_2-r_1}-V_{\text{loc}}\right) ,
	\eeq
	where $(r_1, r_2)$ are the corresponding weights of the two $sl(2,\mathbb{C})$-triplets in the intersection locus of the singular point. The different values than can happen depend on the type of singularities (and enhancement) that is realized in each specific case. Recall that the coefficients $V_{r_1 r_2}$ represent the different components of the $G_4$ flux in the limiting Hodge splitting of $H_p^4 (X_4, \mathbb{R})$ (actually an axion dependent combination thereof), and we have loosen the correlation between them so as to eliminate the positive definiteness constraint and accommodate more generic vacua. Also note that we've introduced a general dependence on the exponent of the dilaton $s$ in the prefactor. This intends to capture the different limits one can approach in the type IIA setting. Henceforth, though, we will stick to the case $\eta=3$, as this is the one corresponding to the weak coupling and large volume limit, where our classical vacua reside. It was shown in \cite{Grimm:2019ixq} that this limit (which is the dual of the type IIB orientifold arising from Sen's weak coupling limit in F-theory) actually corresponds to a singularity enhancement (in the CY$_4$) of the type $ \text{II}_{0,1} \rightarrow \text{V}_{2,2}$. This means that the allowed values of the pairs $(r_1, r_2)$ can be read directly from table 5.4 in \cite{Grimm:2019ixq}. Let us mention also here that the possible UAM fluxes that appear in this particular enhancement can be found in table 5.5 of \cite{Grimm:2019ixq}.
	
	All in all, the most generic asymptotic flux potential we can get at this singularity is of the form
	\beq
	V_ \text{{IIA}} = \frac{1}{s^3} \left( \frac{V_{F_0} t^3}{s} + \frac{V_{F_2}t}{s} + \frac{V_{F_4}}{st} + \frac{V_{F_6}}{st^3} + \frac{V_{h_0}s}{t^3}+\frac{V_{a}s}{t}+V_{g_1}st+V_{g_2}st^3-V_\text{loc}\right) ,
	\label{iiapot}
	\eeq
	where the naming of the different components has been done such that one can identify them with the contribution coming from the R-R fluxes $F_p$, NS-NS flux $h_0$, metric fluxes $a$, and non geometric ones $g_1,g_2$ (cf. eq \eqref{wgen}). Notice that this discussion is not restricted only to the case in which we have just two moduli. Indeed, it is still valid even when our moduli space has more dimensions but we choose some flux configuration such that the vacuum condition fixes some of them to be proportional to others, leaving us with an effective two-moduli problem.
	
	With all this information we can understand using this language the results in our supersymmetric and non-supersymmetric AdS vacua discussed in section \ref{sss:adsex}, and also the Minkowski no-scale vacuum without metric fluxes of \ref{sss:minkex}. We will discuss in the following each of them in turn. Note also that this formalism does not capture our Minkowski no-scale with metric fluxes example in type IIA of section \ref{sss:minkex}. The reason for this is that, as will become clear later, to have a minimum at parametric control within our fixed growth sector, it is crucial that there is at least one flux component that is unbounded and asymptotically massless, as discussed below \eqref{eq:4realsplitting}. However, as was demonstrated in \cite{Grimm:2019ixq} the only ones that fulfill this condition at weak coupling and large volume correspond to the R-R fluxes $F_6$ ($e_0$ in our notation) and $F_4$ ($e$ in our notation). Thus, given that this model contain none of these special fluxes (see eq. \eqref{wiia}), we expect the moduli to be fixed in terms of fluxes constrained by tadpole cancellation and hence bounded. This is indeed what happens. Those examples moreover cannot be made to treat $s$ and $u$ somewhat symmetrically (as is evident from \eqref{vts}) and thus cannot be translated to this new setting.
	
	Before discussing each of these cases it is necessary to comment on the strategy to look for vacua at parametric control. Thus recall that we are looking for vacua obtained by minimizing \eqref{iiapot}) which result in an asymptotic stabilization of $s,t$. For this to be the case (even when taking $s,t \rightarrow \infty$), it is necessary that the different terms in the potential scale asymptotically in the same way. Otherwise some terms will gain more importance than others in the limit and we loose control on the solution. Hence we look for solutions of the form $s\sim \rho^r, t\sim \rho^q$. Thus, each component $V_{r_1 r_2}$ of the potential will scale as follows
	\beq
	V_{r_1 r_2} \sim \rho^{(r_1-3)r+(r_2-r_1)q}
	\eeq
	Note that this does not need to be the case for the class of unbounded fluxes (if any), as they can be made to scale the desired way to contribute as the other terms in the potential. Thus we need to include those pairs $(r_1, r_2)$ and fix a relation between $r, q$ such that every term included in $V_\text{IIA}$ yields the same asymptotic scaling. However, it is important to keep in mind that for the results of the strict asymptotic approximation to apply we need also to be sure that we stay in the growth sector $\mathcal{R}_{s,t}$, as discussed at the beginning of this subsection. This boils down to the necessity of having $r>q$.
	
    \subsubsection*{AdS vacua}
	
	Let us look at the DGKT-CIF model of section \ref{sss:adsex}. By staring at the superpotential \eqref{wcif} we see that this includes an unbounded flux (the one coming from $F_4$) and components coming from $H_3$ flux ($h_0$ in \eqref{iiapot}) and $F_0$ (Roman's mass). Recall that our vacuum conditions, both in the supersymmetric \eqref{vtscfi} and non-supersymmetric (see discussion after \eqref{wolfe}) vacua, fixed the complex structure $u$ to be proportional to $s$. Thus, effectively we have a two-moduli problem, the one presented and discussed in previous subsections, and we can readily apply the results. First of all, we see that for the terms to scale the same way asymptotically and give a minimum at parametric control we need $r=3q$, i.e. $s\sim t^3$, which can be seen to be the case in all of our AdS vacua (see e.g. \eqref{vtscfi}). With this relation, one can check that all the terms appearing in $V_\text{IIA}$ (including $V_\text{loc}$) yield to the same scaling, and also (since $r>q$) we keep ourselves within the fixed growth sector, so that we can trust our minima in the limit. This reinforces the discussion in section \ref{ss:iics}, where it was argued that the potential at the minima, $V_0$, should scale as the contribution from the localised sources $V_\text{loc}$ yielding to a dependence exclusively on the scalars in the underlying quaternionic space (i.e. dilaton and complex structure). Notice that this is valid both for the supersymmetric and non-supersymmetric cases. In the former $V_0=-3 m_{3/2}^2$, and this last statement extends immediately to the gravitino mass. For the non-supersymmetric case even though we don't have this explicit relation between the c.c. and the gravitino mass, its square is still proportional (see eq. \eqref{eq:m32}) to the unique negative summand in the Cremmer et al formula \eqref{eq:VFterm} for the $\mathcal{N}=1$ supergravity potential and hence should have the same scaling as the other summands. 
	
	Recall that we also had AdS extrema by including metric fluxes. This is discussed around \eqref{wcif2}. There we were looking specifically for non-supersymmetric solutions and we specialized to the case were no constant moduli-independent contribution to $W$ was included, i.e. we set $e_0=0$ (see eq. \eqref{wcif2}). We could, in principle, have included such a contribution, and the analysis for the vacua with metric fluxes would have proceed analogously (see \cite{cfi} for more details). In the following, we will consider this latter situation, which simply amounts to change the value of the K\"ahler structure saxion $t$ in the vacuum from \eqref{vevscfi2} to the following one \cite{cfi}
	\beq
	\pm 9ct^2=e_0-\frac{h_0 e}{a}-\frac{h_0^2c}{3a^2}
	\label{tsquare2}
	\eeq
	The vacuum conditions fixed again $s\sim u$, so that we can discuss these vacua within this context as well. Note that to have indeed a solution we had to set $m=0$. This means that we are considering again a flux potential with two unbounded and asymptotically massless fluxes (those coming from $F_4$ and from $F_6$) and some other components which are bounded by the tadpole condition. Moreover, in order to have a solution at parametric control, it is convenient (although not essential in this case) to turn off the contribution of the $h_0$ flux (which can be seen to be consistent with \eqref{tsquare2} above and with the vacuum condition, as discussed after \eqref{wcif2}, leading to both supersymmetric and non-supersymmetric vacua depending on the sign of $e_0$), as this induces a term in the scalar potential \eqref{iiapot} that scales differently than the others (indeed it grows more slowly), and hence gets ``diluted" with respect to those in the asymptotic limit. One can then see that this case is exactly the other unique AdS solution obtainable through \eqref{iiapot}, as was discussed in \cite{Grimm:2019ixq}. There it was shown that one can get an AdS vacuum at parametric control if $r=q$ is satisfied and contributions both from metric fluxes $a$ and the R-R 2-form $F_2$ are included. In this case, every component scales the same way as the local potential $V_\text{loc}$, validating the arguments in section \ref{ss:iics}. The problem with this possibility is that, as was pointed out in \cite{Grimm:2019ixq}, the relation $r=q$ implies that even though we are in the asymptotic limit, we do not approach it by staying in the growth sector $\mathcal{R}_{s,t}$, and hence the analysis is not valid. In our case this is seen by the fact that we cannot stay within the string perturbative regime, as was mentioned after \eqref{eq:scalesads2} (it corresponds to the blue dot in Figure \ref{triangulo}).  
	
    \subsubsection*{Minkowski No-Scale without metric fluxes}
	
	To end this section we recast our second no-scale Minkowski example in section \ref{sss:minkex} in terms of this language. To make contact with the analysis presented here we will further set $c=e=0$ in \eqref{wiia2}, which can be seen to be compatible with the solution. The reason for doing this is that, after this has been done, we see that the potential we get contains contributions from an unbounded flux (the one coming from $F_6$), and two other bounded fluxes $F_0, h_0$. Note that here again we can make $u\sim s$, as the complex structure is not fixed by the potential. This is entirely analogous to the case of AdS vacua without metric fluxes of the last subsection, where in order to have a minimum at parametric control we needed $s\sim t^3$ to be satisfied, so that every term in the potential, including the one from localised sources, scales the same way. This solution is thus seen to be the one presented in section \ref{sss:minkex} (modulo this identification of the complex structure modulus and the restriction imposed on the fluxes).
	
    \subsubsection*{The parameter $\delta$ in the two-moduli limit}

	This formalism also lets us understand why the $\delta$ parameter in the GDC was 7/9 for the supersymmetric and non-supersymmetric AdS examples discussed in sect. \ref{sss:adsex}, and also for the type IIA Minkowski no-scale vacuum of sect. \ref{sss:minkex} when $M_{KK}^x \sim M_{KK}^y \sim M_{KK}$ (see discussion after \eqref{eq:scalesmink2}). The reason for this is that in these models we can relate the naive compactification scale with the gravitino mass in our parametrically controlled vacua, by exploiting the scaling arguments discussed before. Thus, given the KK mass in a generic type IIA CY$_3$ compactification presented in \eqref{msmkk}, which we recover here for completeness
	
	\beq
	M_{KK} \sim \frac{\ms}{(\mathcal{V}_A)^{1/6}} \sim \frac{\mp}{s t^{1/2}} \ , \, 
	\eeq
	and recalling that in the two-moduli case the only possibility to get (perturbative) vacua at parametric control forced us to satisfy $s \sim t^3$ so that $V_\text{loc} \sim s^{-3}$, one thus sees that the KK mass grows as $s^{-7/6}$. Hence, by the arguments explained in the previous section one obtains
	\beq
	M_{KK} \sim m_{3/2}^\delta \Rightarrow \delta=7/9 \ , 
	\eeq
	in agreement with our results (cf. Figure \ref{triangulo}). Let us mention once again that the Minkowski no-scale model with metric fluxes in type IIA \eqref{wiia} did not enter in our discussion in this section. Recall that the reason for this was that both $t$ and $s$ were fixed at a finite value in terms of quantized fluxes constrained by tadpole restrictions (see \eqref{vts}), leaving the (universal) complex structure modulus $U$ undetermined, on which the on-shell $m_{3/2}$ solely depends. Thus, even though one could let $u$ grow to infinity (which is indeed necessary if one wants to maintain string perturbativity), it is not possible to reach the asymptotic regime in the K\"ahler plus dilaton sector, and hence the model cannot be recast within the lines of this section. However, it is interesting to see that this example lead to a different $\delta$ in the GDC, which was $2/3$ if one focuses on the leading KK tower of \eqref{KKxy4}.
	
	\subsection{The GDC, tensionless membranes and tensionless strings}
	\label{ss:strings}

	In this section we explore some possible implications for the breaking of the EFT of having a membrane that becomes tensionless. In particular, we use the unavoidable appearance of tensionless axionic strings in the same limit, in the spirit of \cite{lmmv, lmmv2}, to give an upper bound for the mass scale of the leading tower that arises as the membrane become tensionless. To keep the analysis as general as possible, we will use the formalism of limiting MHS introduced above, but the reader interested in more concrete realizations can check how the results in this section reduce to the ones in section \ref{ss:stringsmembranesIIA} for type IIA toroidal orientifolds. 
	
	First we restrict ourselves to the growth sector defined in \eqref{growthsector}. Within such a growth sector, the fluxes can be split as
	$\Gamma=\bigoplus_{\mathbf{r}}\, \Gamma_\mathbf{r}$, analogously to the decomposition of the real cohomology given in eq. \eqref{eq:realsplitting}. More precisely, we group the different terms in this splitting as \cite{Grimm:2019ixq, lmmv}
	\begin{equation}
	\label{eq:fluxsplitting}
	\Gamma=\Gamma_{\mathrm{light}} \oplus \Gamma_{\mathrm{heavy}} \oplus \Gamma_{\mathrm{rest}}.
	\end{equation}
	The labels denote the behavior of the Hodge norm of the fluxes within each subspace, namely the tension of the membranes dual to the fluxes (and their contribution to the scalar potential)  within $\Gamma_{\mathrm{light}}$ tends to zero, the ones associated to fluxes that belong to $\Gamma_{\mathrm{heavy}}$ diverge, and $\Gamma_{\mathrm{rest}}$ corresponds to the ones whose behavior is path-dependent. In particular, for a CY$_3$ we have
	\begin{equation}
	\begin{aligned}
	\Gamma_{\text {light }} &=\bigoplus_{\mathbf{r}} \Gamma_{\mathbf{r}}, & \text { with } \mathbf{r}=\left\{r_{1}, \ldots, r_{\hn-1} \leq 0, r_{\hn}<0\right\} \\
	\Gamma_{\text {heavy }} &=\bigoplus_{\mathbf{r}} \Gamma_{\mathbf{r}}, & \text { with } \mathbf{r}=\left\{r_{1}, \ldots, r_{\hn-1} \geq 0, r_{\hn}>0\right\} .
	\end{aligned}
	\end{equation}
	
	In the strict asymptotic approximation, the tension of a membrane with charge $\mathbf{q_r}$ in a single $\Gamma_\mathbf{r}$ (i.e. a membrane that interpolates between a fluxless Minkowski region and one with the fluxes given by $\mathbf{q_r}$) takes a particularly simple expression. It can be computed by using the sl(2)-approximation for the K\"ahler potential and the superpotential, which take the form \cite{Grimm:2019ixq} 
	\begin{equation}
	\label{eq:KWstrict}
	\begin{aligned}
	K\, \simeq \, & - \log \left\{ (t^1)^{d_1} \,  (t^2)^{d_2 - d_1} \, \ldots \, (t^{\hn})^{d_{\hn}-d_{\hn -1}} \right\}, \\ 
	W_{\mathbf{q_r}}\, \simeq \, &  \rho_{\mathbf{r}} \left(\mathbf{q_{r}}, b \right) \left( t^{1}\right)^\frac{d_{1}+r_{1}}{2}\left( t^{2}\right)^\frac{d_{2}-d_{1}+r_{2}-r_{1}}{2} \ldots\left( t^{\hat{n}}\right)^\frac{d_{\hn}-d_{\hn-1}+r_{\hn}-r_{\hn-1}}{2}\, ,
	\end{aligned}
	\end{equation}
	and substituting into eq. \eqref{eq:TmemBPS} yields \cite{lmmv}
	\begin{equation}
	\label{eq:Tmemqr}
	T_{\mathbf{q}_{r}} \simeq \, T_{0}\,  \rho_{\mathbf{r}}\left(\mathbf{q_r} , b^{i}\right) (t^{1})^{\frac{{r}_{1}}{2}} (t^{2})^{\frac{{r}_{2}-{r}_{1}}{2}} \ldots (t^{\hn})^{\frac{{r}_{\hn}-{r}_{\hn-1}}{2}},
	\end{equation}
	where $T_0$ has units of $M_P^3$ and it is finite, even though it can depend on the moduli that are not taken to infinity (if any) and $\rho_{\mathbf{r}}\left(\mathbf{q_r} , b^{i}\right)=e^{b^i N_ i} \mathbf{q_r}$ depends on the axions and the fluxes and is also finite.\footnote{In models in which some fluxes are not restricted by a tadpole constraint, there can be an extra divergence associated to the flux becoming large. However, in all the cases we are interested in, the leading contribution to the tension of the membranes can be captured by a finite flux number.}
	
	On the other hand, there exist BPS strings in 4d $\mathcal{N}=1$ EFTs  which implement monodromies $b^j \rightarrow b^j + e^j$ when they are encircled, with $e^i$  the charge of the string. Their tension is given by \cite{lmmv}
	\begin{equation}
	T_{\mathrm{string}}=\left|\frac{1}{2}  e^j \frac{\partial K}{\partial t^j}\right|.
	\end{equation}
	Taking the approximation for the K\"ahler potential given in eq. \eqref{eq:KWstrict} for the strict asymptotic approximation, we see that for each field that goes to infinity there is a string whose leading contribution to the tension takes the form
	\begin{equation}
	\label{eq:Tstringbps}
	T_{\mathrm{string}}^j\simeq \dfrac{d_j - d_{j-1}}{2 t^j}.
	\end{equation}
	Notice that the above expression nicely matches the expressions given in Table \ref{tablaasymp1}. Let us recall that this is in agreement with the idea put forward in \cite{lmmv, lmmv2} that any infinite distance is associated with an axionic string becoming tensionless. However, note that this expression is only valid for the cases in which $d_j - d_{j-1} \neq 0$, signaling that the process of approximating the periods by means of their nilpotent or sl(2)-orbits, as required to obtain eq. \eqref{eq:KWstrict}, does not generally commute with taking derivatives with respect to the moduli (see \cite{Bastian:2020egp, Grimm:2020ouv,Grimm:2021ikg} for more details on this issue). This implies that the approximation in eq. \eqref{eq:KWstrict} does not always include the necessary information to calculate these derivatives, and we should first calculate the derivative of the full K\"ahler potential  and then use the nilpotent or sl(2)-approximation.  However, this is a rather complicated task in general and we will restrict ourselves to the cases in which  $d_1\neq 0$ (i.e. we approach an infinite distance singularity when $t^1 \rightarrow \infty $), as this turns out to be enough for our purposes in this section. Note that this is not a very restrictive assumption as long as we study infinite distance points, and it is enough to compute the tension of the leading string tower in the given growth sector.
	
	Let us now use this information to give a lower bound for $\delta$ in the GDC. As explained in section \ref{ss:wgc}, the gravitino mass  in Planck units, given in eq. \eqref{eq:m32}, takes the same form as the tension of a membrane interpolating between a fluxless Minkowski vacuum and the flux configuration in which we want to study the gravitino mass, displayed in eq. \eqref{eq:TmemBPS}. That is 
	\begin{equation}
	\label{eq:m32strict}
	\dfrac{m_{3/2}}{M_P}=e^{K/2}|W|=\dfrac{1}{2}\dfrac{T_{\mathrm{mem}}}{M_P^3}.
	\end{equation}
	Therefore, sending the gravitino mass to zero while staying in the growth sector defined in eq. \eqref{growthsector} implies that the membrane with charge vector $\mathbf{q}$ equal to the fluxes becomes tensionless. In particular, following \eqref{eq:fluxsplitting} the charge can be decomposed as $\mathbf{q}=\sum_\mathbf{r} \mathbf{q_r}$ and therefore all the constituent membranes with charges $\mathbf{q_r}$, whose tensions are given by eq. \eqref{eq:Tmemqr}, would also be tensionless, implying that all these $\mathbf{q_r} \in \Gamma_{\mathrm{light}}$. 
	We obtain then, for the leading contribution to the tension given by the heaviest $T_{\mathbf{q_r}}$, the following expression
	\begin{equation}
	\label{eq:m32bound}
	\dfrac{m_{3/2}}{M_P}\simeq \dfrac{T_{\mathbf{q_r}}}{M_P^3}= \dfrac{T_0}{M_P^3} 
	\rho_{\mathbf{r}}\left(\mathbf{q_r}, b^{i}\right) 
	\left( t^{1}\right)^\frac{r_{1}}{2}\left( t^{2}\right)^\frac{r_{2}-r_{1}}{2} \ldots\left( t^{\hn}\right)^\frac{r_{\hn}-r_{\hn-1}}{2} \, .
	\end{equation}
	Our goal now is to obtain a lower bound for the $\delta$ in the GDC. First, we define $\hat{r}_i =r_j- r_{j-1}$, with $r_0=0$. Second, we recall that in the given growth sector $t^1>t^j>1$ for all $j>1$, and in order to bound the gravitino mass from below we proceed by following the subsequent steps (where we have omitted the factors of $1/2$ in the exponents in order not to clutter the notation):
	\begin{itemize}
		\item[1)]{For every ${\hat{r}}_j\leq 0$,\footnote{Recall that all $r_j \leq 0$ but this does not necessarily imply ${\hat{r}}_j \leq 0$.} we use $(t^j)^{\hat{r}_j}\geq (t^1)^{\hat{r}_j}$ for all $j$.}
		\item[2)]{For $\hat{r}_j>0$, if it is followed by  $\hat{r}_{j+1}< 0$, we have $(t^j)^{\hat{r}_j} (t^{j+1})^{\hr_{j+1}}=$\\ $=\left(\frac{t^j}{t^{j+1}}\right)^{\hr_j}(t^{j+1})^{r_{j+1}-r_{j-1}}\geq (t^{j+1})^{r_{j+1}-r_{j-1}}$, where we have used that  $\left(\frac{t^j}{t^{j+1}}\right)^{\hr_j}\geq 1$. Now, if $r_{j+1}-r_{j-1}\leq 0$, we can proceed as in step 1) and utilize $(t^{j+1})^{r_{j+1}+r_{j-1}} \geq (t^1)^{r_{j+1}+r_{j-1}}$. On the other hand, if $r_{j+1}-r_{j-1}>0$ we go back again to the beginning of  step 2) if $\hat{r}_{j+2}\leq 0$ or go to step 3 otherwise.}
		
		\item[3)]{If we have $\hr_j>0$ followed by  $\hr_{j+1}> 0$, we use that $(t^j)^{\hr_j} (t^{j+1})^{\hr_{j+1}}\geq (t^{j+1})^{\hr_j+\hr_{j+1}}=(t^{j+1})^{r_{j+1}-r_{j-1}}$. If the exponent $\hr_{j+2}\leq 0$ we go to step 2) and if it is $\hr_{j+2}\geq 0$ we proceed as in step 3) again. }
	\end{itemize}
	After following these steps systematically for all $t^j$ we arrive at
	\begin{equation}
	\dfrac{m_{3/2}}{M_P}\geq \dfrac{T_{\mathbf{q_r}}}{M_P^3}\geq \dfrac{T_0\, \rho_{\mathbf{r}} }{M_P^3} \ (t^1)^{\frac{r_{\mathrm{min}}}{2}}
	\end{equation}
	where $r_{\mathrm{min}}=\mathrm{min}(r_j)$. Using eq. \eqref{eq:Tstringbps} we can relate $t^1$ to the tension of the corresponding BPS string as long as we are studying infinite distance singularities (i.e. $d_1>0$), obtaining
	\begin{equation}
	\label{eq:m32deltacrit}
	\left(\dfrac{m_{3/2}}{M_P}\right)^{\frac{1}{\left| r_{\mathrm{min}}\right| }}\gtrsim \dfrac{(T_\mathrm{string}^1)^{1/2}}{M_P} \simeq \dfrac{m_{\mathrm{tower}}}{M_P}\, ,
	\end{equation}
	where $\gtrsim$ means that we are neglecting finite factors like $T_0$, $\rho_{\mathbf{r}}$ or $d_1$, which are positive but not relevant for the asymptotic behavior, and we have used that the mass of the states associated to the string tower scales as $m_{\mathrm{tower}}^2 \simeq T_{\mathrm{string}}$. We can therefore give a lower bound for the $\delta$ in the GDC because there is always a tower of states associated to a BPS string becoming tensionless whose mass is lighter or equal than $m_{3/2}^{\delta_\mathrm{crit}}$. Consequently, we obtain that for the GDC
	\begin{equation}
	\label{eq:generalbounddelta}
	1 \, \geq \, \delta \, \geq \, \delta_{\text{crit}} \, = \, \dfrac{1}{|r_{\text{min}}|} \, . 
	\end{equation}
	For a CY$_3$, the minimum value for $r_{\mathrm{min}}=-3$ and it can only be obtained if the enhancement chain reaches a type IV singularity, whereas the minimum value for a CY$_4$ may be obtained at a type V singularity and is $r_{\mathrm{min}}=-4$. Thus, we can give the following lower bounds for the parameter in the GDC
	\begin{equation}
	\label{eq:deltamin}
	\begin{aligned}
	\delta \, \geq  & \,\dfrac{1}{3} \qquad \text{for CY$_3$,}\\
	\delta \,  \geq &  \, \dfrac{1}{4} \qquad \text{for CY$_4$.}
	\end{aligned}
	\end{equation}

	Strictly speaking, when applied to type II theories compactified on CY$_3$, this analysis only captures the complex structure sector of type IIB and (by mirror symmetry) the K\"ahler sector of type IIA. However, as seen in previous sections, for the type IIA setup it is not possible to stay within the perturbative regime when we approach the Large Volume Point if this is not accompanied by an infinite displacement in the complex structure sector. One can try to accommodate this by relaxing the form of the K\"ahler potential in eq. \eqref{eq:KWstrict} and using the monodromy generators associated to the complex structure sector of type IIA, as defined in \cite{HIMZ, fhi}. We will see momentarily that this allows us to recover the results from previous sections as well as to recover the adequate F-theory results in the right approximation. First, following \cite{lmmv}, we allow for the the K\"ahler potential to be expressed as 
	\begin{equation}
	\label{eq:Kstrictn}
	K\, \simeq \,  - \log \left\{ (t^1)^{n_1} \,  (t^2)^{n_2 - n_1} \, \ldots \, (t^{\hn})^{n_{\hn}-n_{\hn -1}} \right\}, 
	\end{equation}
	with $n_j$ not necessarily equal to the $d_j$ defined by the action of the associated monodromy generator on the period vector, as in eq. \eqref{eq:defd_j}.\footnote{This is the case when the K\"ahler potential cannot be written directly as $K=-\log \left( \Pi^I \eta_{IJ} \bar{\Pi}^J \right)$.} Thus eqs. \eqref{eq:Tmemqr} and \eqref{eq:m32bound} take the same form with the replacement $r_j \rightarrow r_j^\prime =r_j+ (d_j- n_j)$. 	For type IIA orientifolds, the no-scale condition for the complex structure moduli space yields a maximum value for $n_j=4$. Furthermore, we also have $d_j=1$ because the associated log-monodromy matrix has nilpotency order equal to 1 (see \cite{HIMZ, fhi} for the concrete expressions). The question now is whether this may give a different value for $\delta_{\mathrm{crit}}$. From the $(d_j-n_j)$ part, the minimum value that can be obtained is 3. Moreover, if we consider a $\mathbf{q_r}$ that does not include NS-NS flux (i.e. $r_j=-1$, as can be seen from the superpotential in eq. \eqref{eq:KWstrict}) we obtain a $\delta_{\mathrm{crit}}=1/4$. Note, however, that this does not mean that we have an example which saturates this bound. In fact, we recover the value $\delta_{\mathrm{crit}}=1/3$ as soon as we consider the presence of NS-NS fluxes (which are typically necessary for cancelling the tadpole sourced by the orientifold), as their contribution to the superpotential (linear on the corresponding complex structure moduli) corresponds to the value $r_j=+1$, yielding $r_j^\prime=-2$, which would not modify the argument below \eqref{eq:m32deltacrit}.
	
	Moreover, the bound $\delta_{\mathrm{crit}}=1/4$ can be  interpreted in a more natural way as the lower bound that one would obtain from an F-theory compactification on a CY$_4$, as the complex structure deformations on such a setup capture the behavior of the dilaton in the dual type IIA compactifications, which belongs to the complex structure moduli space in this case and in turn captures its behavior in the limit $s\sim u^i$.
	
	Finally, it is remarkable that the lower bound $\delta \geq 1/3$ for CY$_3$, also found in section \ref{ss:limits} from different arguments, matches the lower bound for the parameter $\alpha\geq \sqrt{1/6}$ in the SDC found in \cite{Gendler:2020dfp, Andriot:2020lea, Bastian:2020egp} as explained in section \ref{ss:SDCandGravitino}.

	\subsection{Generalizing the relation between the GDC and the SDC}
	\label{ss:SDCandGravitino2}
	
	Recall that in section \ref{ss:SDCandGravitino} a relation between the GDC and the SDC was pointed out, at least for a class of Minkowski no-scale vacua, where some of the moduli remained undetermined and hence there was still some freedom to move within the unfixed field space. There it was shown that if we chose to move along those directions where the gravitino was rendered asymptotically massless, one could relate the relevant exponents in both conjectures to each other, by identifying the tower of light states whose appearance signals the breakdown of the EFT. Indeed, by particularizing to the type II examples analysed in section \ref{ss:iiaex}, we obtained a relation of the form $\alpha = \sqrt{\frac{3}{2}}\,  \delta \ $, which was shown to reproduce the known bounds in the $\alpha$ parameter of the SDC existing in the literature \cite{Gendler:2020dfp, Andriot:2020lea, Bastian:2020egp}. Henceforth our goal will be to generalize this discussion by using the tools of Asymptotic Hodge Theory introduced in this section. 
	
	To begin with, recall that the general relation between the $\alpha$ parameter in the SDC and the $\delta$ in the GDC along a given geodesic path towards a singularity is given in eq. \eqref{eq:alphadelta}, that we recall here for definiteness
	\begin{equation}
	\label{eq:alphadeltabis}
	\dfrac{\alpha}{\delta}=-\dfrac{\log (\m32)}{d} \, .
	\end{equation}
	First, let us try to calculate the geodesic distance $d$. Consider the sl(2)-orbit approximation of the K\"ahler potential \eqref{eq:KWstrict}, which extracts the leading dependence on the moduli that diverge (after restricting to a specific growth sector of the form \eqref{growthsector}) yielding the following expression in terms of the complex moduli $T^i$:\footnote{Here we keep the notation used in the rest of section \ref{ss:MHS} above, denoting the divergent moduli in general as $T^i$. To make contact with the results in section \ref{ss:SDCandGravitino}, recall that the divergent moduli there are the complex structure moduli, denoted $U^i$.}
	\beq
	K \simeq - \text{log} \left( (T^1+\bar{T}^1)^{d_1} (T^2+\bar{T}^2)^{d_2-d_1}\ldots \ \right),
	\eeq
	where the $d_i$ were defined in \eqref{eq:defd_j} and indicate the singularity type at each step along the enhancement chain (see discussion around \eqref{chain}).\footnote{To allow for more general K\"ahler potentials one could make the replacement $d_ i\rightarrow n_i$ as explained around eq. \eqref{eq:Kstrictn}.} Note that this implies that the K\"ahler metric in the subsector of the divergent moduli is that of a direct product of hyperbolic planes \cite{Calderon-Infante:2020dhm}, \footnote{Note that this expression is only valid whenever $d_i - d_{i-1}\neq 0$. If this is not the case, one could first take the derivatives and then use the approximation given by the sl(2)-orbit, but this is extremely more involved and beyond the scope of this work, so we limit our discussion here to the cases in which $d_i - d_{i-1}\neq 0$ (see \cite{Bastian:2020egp,Grimm:2020cda, Grimm:2021ikg}).} 
	\beq
	K_{T^i \bar{T}^i}=\frac{d_i-d_{i-1}}{4 (t^i)^2} ~~ \text{and} ~~ K_{T^i \bar{T}^j}=0 ~~ \text{for} ~~ i \neq j.
	\eeq
	
	By solving the geodesic equation with the metric given by $ds^2\, =\,  2 \, K_{T^i \bar{T}^j} \, dT^i \, d\bar{T}^j$ in such a limit one can see that the unique geodesic curves that asymptote to $t^i= \text{Re}\ T^i \to \infty$ are those for which the axions are constant and the rate at which each of the saxions grows with respect to the other is not fixed. More precisely, all such geodesics towards the singularity at infinity can be expressed as
	\begin{equation}
	t^1\, = \, \gamma_{2} \cdot (t^2)^{\mu_2}\, =\, \ldots \, = \, \gamma_{\hn} \cdot (t^{\hn})^{\mu_{\hn}},
	\end{equation}
	where $\gamma_i$ are finite constants that only depend on the initial points and become irrelevant as we approach the singularity, and the exponents fulfill $\mu_i \geq \mu_{i+1}$ with $1\geq \mu_i\geq 0$, so that the trajectory stays within the chosen growth sector as the singularity is approached.\footnote{This restriction on the exponents does not imply any loss of generality since any other geodesic not fulfilling it can be accommodated within a different growth sector with analogous restrictions in the exponents.} Along such geodesics, the distance can be expressed as follows
	\begin{equation}
	\label{eq:geodesicd}
	d=\sqrt{\dfrac{d_1}{2} \log^2  (t^1)+ \dfrac{d_2-d_1}{2} \log^2 (t^2) + \ldots +\dfrac{d_{\hn}-d_{\hn-1}}{2} \log^2 (t^{\hn})}.
	\end{equation}
	
	This geodesic distance can then be bounded both from above and from below as
	\begin{equation}
	\label{eq:boundsd}
	\sqrt{\dfrac{d_{\hn}}{2}} \, \log (t^{\hn})  \,  \geq\,  d \, \geq \,  \sqrt{\dfrac{d_1}{2}} \,  \log (t^1) \, .
	\end{equation}
	For the lower bound we have used the fact that all the terms in the sum of eq. \eqref{eq:geodesicd} are non-negative, and for the upper bound the fact that the contribution from each of the terms in the sum is maximized when all the $\mu_i \rightarrow 1$ and therefore $t^1\simeq t^2 \simeq \ldots \simeq t^{\hn}$ (neglecting the contribution from the $\gamma_i$, which can be used for the path to remain within the desired growth sector, but whose contribution to the distance becomes negligible in the limit $t^i \rightarrow \infty$). Finally, let us remark that the reason why we have chosen to express the upper bound in terms of $t^{\hn}$ and the lower bound in terms of  $t^1$ will become clear shortly.
	
	Given these bounds for the distance, we turn now to examining the piece depending on $\log (\m32 )$. First of all, note that as a consequence of the logarithm, any constant prefactor appearing in the gravitino mass, as well as any subleading additive contributions, may be neglected in the limit of vanishing gravitino mass, so we only need to keep track of the dependence on the divergent moduli. It can therefore be bounded both from above and from below as
	\begin{equation}
	\dfrac{r_{\text{min}}}{2} \, \log (t^{\hn}) \, \geq \, \log (\m32 ) \geq \,   \dfrac{r_{\text{min}}}{2} \, \log (t^1)  \,,  
	\end{equation}
	where we recall that $r_{\text{min}}\leq -1$ and it has the highest absolute value of all $r_i$ that enter the gravitino mass. The lower bound is obtained directly from the eq. \eqref{eq:m32deltacrit}, and the upper bound by applying a reasoning analogous to the one explained in the three steps described before eq. \eqref{eq:m32deltacrit} but going from $t^n$ to $t^1$ and bounding everything from above.
	
	When combined with eq. \eqref{eq:alphadeltabis}, these bounds can be used to restrict the ratio between the parameters in the SDC and the GDC, yielding
	\begin{equation}
	\label{eq:boundsalphadelta}
	\dfrac{|r_{\text{min}}|}{\sqrt{2 d_1}} \, \geq \,  \dfrac{\alpha}{\delta} \, \geq \,  \dfrac{|r_{\text{min}}|}{\sqrt{2 d_{\hn}}}
	\end{equation}
	where it becomes clear that expressing the previous bounds for the distance and the logarithm of the gravitino mass in terms of $t^1$ and $t^{\hn}$ was the right choice for the dependence on the moduli to drop from this bound.
	
	Let us now remark a few interesting points. First of all, in the one-modulus case the lower and the upper bounds degenerate (as expected from the fact that there is only one geodesic since the space is one-dimensional). This would yield  $\alpha\, =\, \frac{|r_{\text{min}}|}{\sqrt{2 d_{\hn}}} \, \delta$, and its highest value for a CY$_3$ would be obtained at a type IV singularity giving $\alpha\, =\, \delta \, \sqrt{3/2}$. In fact, this is also the relation obtained in the type IIA setup considered in section \ref{ss:SDCandGravitino} (which is actually dual to a type IIB setup in which a type IV singularity is approached). In order to recover this result from the general formula one needs to make the appropriate replacements to describe the part of field space that is divergent, namely the complex structure moduli in IIA. These replacements, already introduced throughout this section, are  $d_{\hn}\rightarrow n_{\hn}$ and $r_{\hn}\rightarrow r_{\hn}^\prime= r_{\hn}+(d_{\hn}-n_{\hn})$. Then, the corresponding values for the setup of section \ref{ss:SDCandGravitino}, which can be read from the K\"ahler potential and the superpotential, turn out to be $r_{\hn}=-1$, $d_{\hn}=1$ and $n_{\hn}=3$, so that one recovers the result $\alpha\, =\, \delta \, \sqrt{3/2}$ as expected.
	
	Second, one can combine the bounds in eq. \eqref{eq:boundsalphadelta} with the ones for $\delta$ displayed in eq. \eqref{eq:generalbounddelta}, obtaining the following bounds for the SDC parameter
	\begin{equation}
	\label{eq:boundsalpha}
	\dfrac{|r_{\text{min}}|}{\sqrt{2 d_1}} \, \geq \, \alpha \, \geq \,  \dfrac{1}{\sqrt{2 d_{\hn}}}.
	\end{equation}
	Moreover, for a CY$_3$ this implies $\frac{3}{\sqrt{2}} \, \geq \, \alpha \, \geq \frac{1}{\sqrt{6}}$, which coincides with the lower bounds for $\alpha$ found in \cite{Andriot:2020lea, Gendler:2020dfp,Bastian:2020egp}.

	\section{Phenomenological implications}
	\label{sec:pheno}
	
	The value of the gravitino mass is an important phenomenological parameter both in particle physics and cosmology.
	In Minkowski vacua the gravitino mass is a direct measure of the scale of supersymmetry breaking. 
	But also for large moduli  the gravitino mass gives the size of supersymmetry breaking in runaway dS and AdS minima.
	In supersymmetric theories of particle physics the gravitino mass will typically give us the scale of the mass of the supersymmetric 
	partners of the SM.  If the GDC introduced in this paper is true, an important message is
	that {\it one cannot arbitrarily decouple the gravitino mass from the UV scales}. 
	One rather has	
	\beq
	m_{3/2}\ \simeq \ \frac {M^{1/\delta}_{KK}}{M^{(1-\delta)/\delta}_{\text{P}}} \   \ ,\ \ \delta \ <\ 1 \ .
	\label{limites}
	\eeq
	For $\delta=1$ there is no decoupling and the gravitino mass would be of order the KK scale. 
	The maximal scale separation is reached for the smallest possible $\delta$. We have seen in the previous sections that 
	in general the minimal $\delta$ is $\delta=1/3$, although in the specific examples analysed only $\delta=2/3$ for the lightest
	tower is reached. Thus e.g. for such a value one would have the maximum separation at
	\beq
	m^2_{3/2}\ \simeq \  \frac {M^3_{KK}}{\mp} \ .
	\eeq
	The implications for particle physics depend on the actual value of $m_{3/2}$.  A couple of phenomenologically interesting
	values for the gravitino mass are:
	\begin{itemize}
		\item[1)] \ $m_{3/2}\sim 1	\ \text{TeV}$.  This is the popular case in low-energy  MSSM supergravity models, in which the gravitino 
		mass is tied to the electro-weak scale. Then eq.(\ref{limites}) implies an upper limit 
		\beq
		M_{KK}\ \lesssim \ 10^{8}\ -\ 10^{13} \ \text{GeV}
		\eeq
		for $\delta = 2/3\ -\ 1/3$. Thus the {\it big desert scenario} with no new physics beyond the MSSM up to a scale
		$M_X\sim 10^{16}$ GeV would not be consistent with the GDC.
		\item[2)] \ $m_{3/2}\sim 10^{10}	\ \text{GeV}$.  This is the intermediate scale scenario in which the 
		non-supersymmetric SM
		is valid up to an intermediate scale $10^{10}$ GeV \cite{Hall:2009nd,Ibanez:2013gf}. It is well known that above those energies the SM scalar potential
		with a Higgs mass $m_H=125$ GeV  becomes  unbounded from below.  The virtue of having $m_{3/2}$  at this intermediate scale is that
		then supersymmetry is restored and the potential becomes stable and positive. In this case the UV scales may be a bit larger with
		\beq
		M_{KK}\ \lesssim \ 10^{13}\ -\ 10^{16} \ \text{GeV}
		\eeq	 
		for $\delta = 2/3\ -\ 1/3$. 
	\end{itemize}
	It is interesting to note that a gravitino mass lower bound in terms of membrane tensions was already 
	remarked in section 5 of ref. \cite{Herraez:2016dxn}. There the necessity of a lower UV scale in order to
	obey the WGC as applied to membranes was pointed out.
	
	From the point of view of cosmology,  the difficulties to accommodate inflation in a way consistent with the 
	dS swampland conjecture are well known. From the GDC here studied an obvious condition is that the 
	Hubble constant upon inflation must be $H\lesssim M_{KK}$, for an effective field theory description to make 
	sense. Thus one can write
	\beq
	H\ \lesssim \ m_{3/2}^\delta \ M^{(1-\delta)}_{\text{P}} \ .
	\eeq
	For the range of gravitino values  mentioned above one thus  generically expects negligible tensor 
	perturbations in the cosmic background. Recently it has been pointed out that one should impose as a
	Swampland condition that the cosmological gravitino sound speed should be non-vanishing
	\cite{Kolb:2021nob,Kolb:2021xfn}. It would be interesting to study whether this condition and
	our GDC are consistent.

	It would also be interesting to study whether the GDC is consistent with well-known 
	scenarios to fix all moduli in dS in  type IIB string theory like KKLT \cite{Kachru:2003aw}
	or the LVS \cite{Balasubramanian:2005zx}.
	Since both scenarios start 
	with AdS vacua a first question is whether they are consistent with the ADC.
	In ref.\cite{Blumenhagen:2020dea} it is claimed that the  AdS step of  KKLT is consistent with the ADC under certain circumstances.
	In particular, if the
	KKLT requirement $|W_0|\ll 1$ is obtained via a strongly warped throat close to a conifold point
	(this is the throat where the ${\overline{\text D3}}$'s would be located),
	then it is claimed  that there is a KK tower associated to this singularity and that its scale is
	\beq
	M^2_{KK} \ \simeq \frac {1}{\log^2(-\Lambda)} |\Lambda|^{1/3} \ ,
	\eeq
	so that the ADC is respected. Since this is  an AdS ${\cal N}=1$ vacuum,
	this would be consistent with the GDC with $\delta=1/3$.

	\section{Final comments and conclusions}
	\label{sec:conclusions}
	
	Many tests in the Swampland program are performed by studying the structure of specific string vacua 
	in some large moduli direction. That is in fact the case of the Swampland Distance Conjecture and the
	AdS Distance Conjecture. In this paper we have emphasized that there is a particularly relevant limit 
	which involves a physical particle rather than a random field direction. This is the limit in which the
	gravitino mass goes to zero, $m_{3/2}\rightarrow 0$. This limit selects particular field directions and
	makes contact with both the Swampland Distance Conjecture and the AdS Distance Conjecture. We propose a Gravitino Distance Conjecture
	which states that in that limit, an infinite tower of massless particles appear, with masses controlled by the gravitino mass
	as $\Mt \sim m_{3/2}^\delta$ (in Planck units), where $\delta$ is a positive constant. The lowest lying tower 
	is typically a KK tower with subleading towers coming from tensionless strings. There is also a direct connection between 
	a vanishing gravitino mass and membranes becoming tensionless. Their  corresponding gauge couplings to 
	3-forms go to zero as $m_{3/2}\rightarrow 0$, which would violate Weak Gravity Conjecture arguments, and hence gives 
	support to the singular character of that limit.
	
	We have presented evidence for this conjecture within the context of  type IIA CY$_3$ orientifold vacua, both by studying specific 
	classes of type IIA toroidal models   as well as considering general properties within type II CY$_3$ and F-theory CY$_4$ settings, using 
	the MHS formalism.
	The exponent $\delta$ is bounded  as $1> \delta \geq 1/3$ for CY$_3$ orientifolds (although $\delta\geq 2/3$ is realized in toroidal examples)
	and $1 > \delta \geq 1/4 $ in F-theory compactifications on CY$_4$. The  Gravitino Distance Conjecture here proposed implies the AdS Distance Conjecture and the exponents
	for the latter are simply given by $\delta/2$.  The value of $\delta$ is also directly connected to the corresponding exponent $\alpha$ 
	in the Swampland Distance Conjecture, and it is shown that  $\frac{\alpha}{\delta}=\sqrt{\frac {3}{2}}$ in toroidal orientifolds. In general CY$_3$ compactifications we recover the same result for the one-modulus case and for several moduli we obtain $\frac{3}{\sqrt{2}}\geq \frac{\alpha}{\delta} \geq \frac{1}{\sqrt{6}}$.
	
	The $m_{3/2}\rightarrow 0$ limit considered by the Gravitino Distance Conjecture is also relevant from the phenomenological point of view. 
	In general, $m_{3/2}$ sets the scale of supersymmetry breaking  (except for supersymmetric AdS, which is not phenomenologically relevant).
	If one wants to have a low gravitino mass in order to address the hierarchy problem ($m_{3/2}\sim 1$ TeV) or to
	guarantee the stability of the Higgs potential in the Standard Model ($m_{3/2}\lesssim 10^{10}$ GeV),  the GDC tell us that this comes
	along with a relatively low KK scale $M_{KK}\sim 10^8-10^{13}$ GeV. This implies that the traditional desert  scenario with a gauge coupling unification at
	$\sim 10^{16}$ GeV  would not be viable.  This lowering of the UV scale also puts limits on a possible inflationary potential, which should fulfill $H\lesssim m_{3/2}^\delta M^{(1-\delta)}_{\text{P}}$ GeV.
	
	There remain many open questions to analyse. In particular, all examples provided are classical type II vacua. 
	This is also the case of the specific tests provided in the literature for the Swampland Distance Conjecture, the AdS Distance Conjecture and even the Weak Gravity Conjecture.
	However the distinction between classical and quantum vacua in string theory is not clear, and it is difficult to believe that all these constraints 
	are only valid for classical vacua. Still it would be important to extend the checks to larger classes of string vacua. It would also be important to
	check whether the Gravitino Distance Conjecture is consistent with specific scenarios for full  moduli stabilisation in dS vacua 
	such as the KKLT or LVS schemes.

	\newpage
	\centerline{\bf Acknowledgments}
	\vspace{.5cm}	
	
	We are grateful to J. Calder\'on, E. Gonzalo, M. Gra\~na,  D. L\"ust, M. Montero, E. Palti, A. Retolaza, M. Scalisi,  C. Vafa, I. Valenzuela, and M. Wiesner for useful discussions. 
	A.F. thanks the IFT UAM-CSIC for hospitality and support at the stages of this work before the pandemic.	
	This work is supported  by  the  Spanish  Research  Agency  (Agencia  Espa\~nola  de  Investigaci\'on) through  the  grants  IFT  Centro  de  Excelencia  Severo  Ochoa  SEV-2016-0597, the grant GC2018-095976-B-C21 from MCIU/AEI/FEDER, UE and the grant PA2016-78645-P. The work of A.C. is supported by the Spanish FPI grant No. PRE2019-089790. The work of A.H. is supported by the ERC Consolidator Grant 772408-Stringlandscape.

	
	\bibliographystyle{JHEP}
	\bibliography{refs}	
	
\end{document}